# Spatiotemporal Imaging of Thermally Driven Metal–Insulator Phase Domain Propagation in $VO_2$

Ilaria Delbono[1,2], Raouf Amara[3,4], Matias Feldman[1], Hankun Yang[3], Cédric Leau[1], Carlo Scian[2], Erwan Dandeu[1], Sébastien Royer[1], Debora Pierucci[1], Emmanuel Lhuillier[1], Franck Vidal[1], Tiziana Cesca[2], Hilton B. de Aguiar[3], Jose Ordonez-Miranda[1], Danièle Fournier[1], Raj Pandya[4,5], Boris Kalinic[2]*, Giovanni Mattei[2]*, James K. Utterback[1]*

[1] Sorbonne Université, CNRS, Institut des NanoSciences de Paris, 75005 Paris, France
[2] Physics and Astronomy Department, University of Padova, via Marzolo 8, I-35131 Padova, Italy
[3] Laboratoire Kastler Brossel, ENS-Université PSL, CNRS, Sorbonne Université, Collège de France, Paris, France
[4] Department of Chemistry, University of Warwick, Coventry, UK
[5] Yusuf Hamied Department of Chemistry, University of Cambridge, Cambridge CB2 1EG, United Kingdom

*E-mail: boris.kalinic@unipd.it
*E-mail: giovanni.mattei@unipd.it
*E-mail: james.utterback@sorbonne-universite.fr

**ABSTRACT**

Achieving a detailed picture of the nature and dynamics of phase boundaries remains a major challenge, particularly solid–solid transitions central to phase-change materials. Here, we use stroboscopic optical scattering microscopy and finite-element simulations to study the nonlinear thermally driven motion of the mesoscale insulator–metal phase boundary in $VO_2$ thin films. We visualize the lateral spatiotemporal evolution of a diffraction-limited photoinduced metallic phase domain, resolving its outward propagation, contraction, and ultimate disappearance. The phase boundary dynamics are governed by heat flow, which can be controlled by excitation fluence, film thickness, and substrate thermal conductivity. Parameter-free agreement between the data and nonlinear thermal transport simulations identify the phase boundary transport as locally diffusive under this regime. We quantify how the latent heat acts as a thermal inertia, suppressing transport and prolonging the phase coexistence. Paradoxically, despite being driven by the underlying thermal gradient locally described by Fourier's law, the mesoscale phase front propagates faster than expected from thermal diffusion. This phenomenon arises from the distinction between temperature-field evolution and phase-boundary isotherm motion. These results establish a link between heat dissipation and mesoscale phase-front motion, providing a framework for controlling phase-transition kinetics relevant to photonics applications of $VO_2$. We anticipate that this approach generalizes to a broad class of first- and second-order transitions.

The insulator–metal transition of correlated materials has been a long-standing focus in condensed matter physics.[1,2] Vanadium dioxide ($VO_2$) stands out as a prototypical solid–solid phase change material for investigating persisting fundamental questions and for emerging phase-change applications.[3–5] Its reversible insulator-to-metal phase transition can be triggered in various ways, including by heating to a moderate temperature of about 341 K,[6] and by optical[7–9] or electrical excitation.[10,11] This transition entails a structural change between the low-temperature insulating monoclinic phase and high-temperature metallic rutile phase, accompanied by a substantial reconfiguration of electronic structure that induces a drastic change in electrical conductivity[12] and optical properties.[13] This is a first-order phase transition, involving latent heat and phase coexistence separated by a phase boundary.

There remains a fundamental debate about the role of nucleation and growth of phase domains, particularly concerning the dynamics of metal–insulator phase-boundary propagation in $VO_2$ thin films.[14] Transition temperature broadening and hysteresis in heterogeneous polycrystalline thin films, and reports of metastable states, further complicate the picture.[14–17] Yet, a better understanding is important because the phase boundary motion dictates the speed, reproducibility, and efficiency of devices based on the insulator–metal transition.[15,16] A complete picture for phase domain dynamics should address the emergent mesoscale behavior across multiple grains of widely used polycrystalline films.[18] It should also provide a physical account of the transport regime governing phase-front motion, including the role of latent heat.[19]

The challenge in directly probing phase-front dynamics lies in the need to generate a controlled phase boundary and track its evolution with high spatio-temporal resolution. Numerous pump–probe methods have extensively investigated the ultrafast insulator-to-metal transition of $VO_2$ over the past two decades.[7,8,20–26] However, typical experiments only provide macroscopically averaged information, leaving the treatment of mesoscale phase domain nucleation, growth, and relaxation to indirect kinetics assignments or simulations.[14,27–29] Time-resolved X-ray imaging has investigated the ultrafast phase change induced by uniform illumination, observing no domain growth or shrinkage in one case,[24] while in another case revealing a displacive lattice transformation following local nucleation under superheating.[18] High-resolution imaging (electron and scanning near-field infrared microscopy) of phase domains has resolved steady-state and ultrafast heterogeneity and hysteresis within individual nanodomains and across nanocrystalline morphologies.[17,30–32] While certain experiments have yielded valuable estimates of phase domain growth rates,[17,18,29,31] the underlying transport regime governing front motion and the role of latent heat remain unclear. Moreover, the reverse process—metal-to-insulator relaxation—remains less explored due to the lack of a direct trigger to induce the transition and its slower time scale. Although a few studies have resolved this relaxation,[19,33–36] the spatiotemporal role of phase domains remains unclear.

Here, we use spatiotemporally-resolved optical pump–probe measurements to access thermally driven metal–insulator phase-boundary propagation and relaxation in $VO_2$ thin films. By stimulating the transition with a diffraction-limited light pulse, we generate a controlled in-plane gradient in the resulting phase change and temperature profile, enabling direct tracking of the mesoscale phase front position as a function of time. We image the metallic domain over a time window of 100 ps–1 μs. Fitting the nonlinear reflectance profile to an optical scattering model monitors the spatial motion of the multi-grain insulator–metal phase boundary with few-nm uncertainty. The kinetics of the metal-to-insulator recovery are consistent with heat dissipation from the $VO_2$ thin film to the substrate. The recovery time scale can be tuned from tens of nanoseconds down to a few nanoseconds by varying the excitation fluence, film thickness, and substrate thermal conductivity. Thermal relaxation drives the in-plane phase boundary motion. We observe that the phase front propagation is time-dependent: the boundary of the thermally sustained metallic domain moves slowly during the first several nanoseconds before rapidly shrinking and disappearing. Finite element simulations of nonlinear, latent-heat-modulated thermal transport reproduce the data well without resorting to adjustable parameters, suggesting that recovery of the insulating phase follows the cooling isotherm as heat flows both laterally (in-plane) and into the substrate (out-of-plane). For time scales beyond electron–phonon equilibration, the metal–insulator phase boundary is thus spatiotemporally driven by thermal gradients. In this framework, the phase-front motion evolves continuously, but tends towards a diffusive regime—behavior qualitatively distinct from a constant-velocity assumption. The calculated temperature profiles reveal nonlinear thermal transport and a plateau at the transition temperature. The model further quantifies how the latent heat of the phase transition slows down the phase propagation. Yet, importantly, the nonlinear temperature evolution allows boundary transport faster than the thermal diffusion associated with the boundary temperature. This latter phenomenon arises naturally without a breakdown in Fourier's law, originating from the distinction between the thermal diffusivity that governs temperature profile broadening and the phase boundary that tracks a fixed isotherm along that profile. These nonlinear dynamics provide fundamental insights and have important implications for photonics applications of $VO_2$.

**$VO_2$ optical properties across phase transition**

The photoinduced phase transition and relaxation were examined in 45 and 180 nm-thick $VO_2$ thin films grown by reactive magnetron sputtering on silica or silicon substrates to tune thermal relaxation (see Supplementary Information Section 1). The samples are polycrystalline with an average grain size of 50 ± 5 nm, as seen by scanning electron microscopy (Fig. 1a). Atomic force microscopy measurements reveal a surface roughness of 3 ± 1 nm (Fig. S1), while grazing-incidence X-ray diffraction patterns confirm the high crystalline quality of the samples in both phases (Fig. S1). The phase transition temperature ($T_{\mathrm{IMT}}$) was determined by the hysteresis cycle of transmittance for both the heating ($T_{\mathrm{IMT,h}}$) and the cooling ($T_{\mathrm{IMT,c}}$) processes (Fig. S2). The real ($n$) and imaginary ($k$) components of the complex refractive index were obtained by ellipsometry, using Bruggeman effective medium theory, at different temperatures to capture the characteristic nonlinear evolution of the optical response across the phase transition (Fig. 1b). The experimentally determined optical constants were employed to calculate the reflectance spectrum (see Supplementary Information Section 1) as a function of temperature for the three investigated samples (Fig. S3). In the following pump–probe measurements, a probe wavelength of 775 nm was used to image the spatiotemporal dynamics. Fig. 1c shows the sigmoidal temperature-dependent variations of the optical constants ($\Delta n$, $\Delta k$) and the resulting change in reflectance ($\Delta R/R_0$) at this wavelength for increasing temperature.

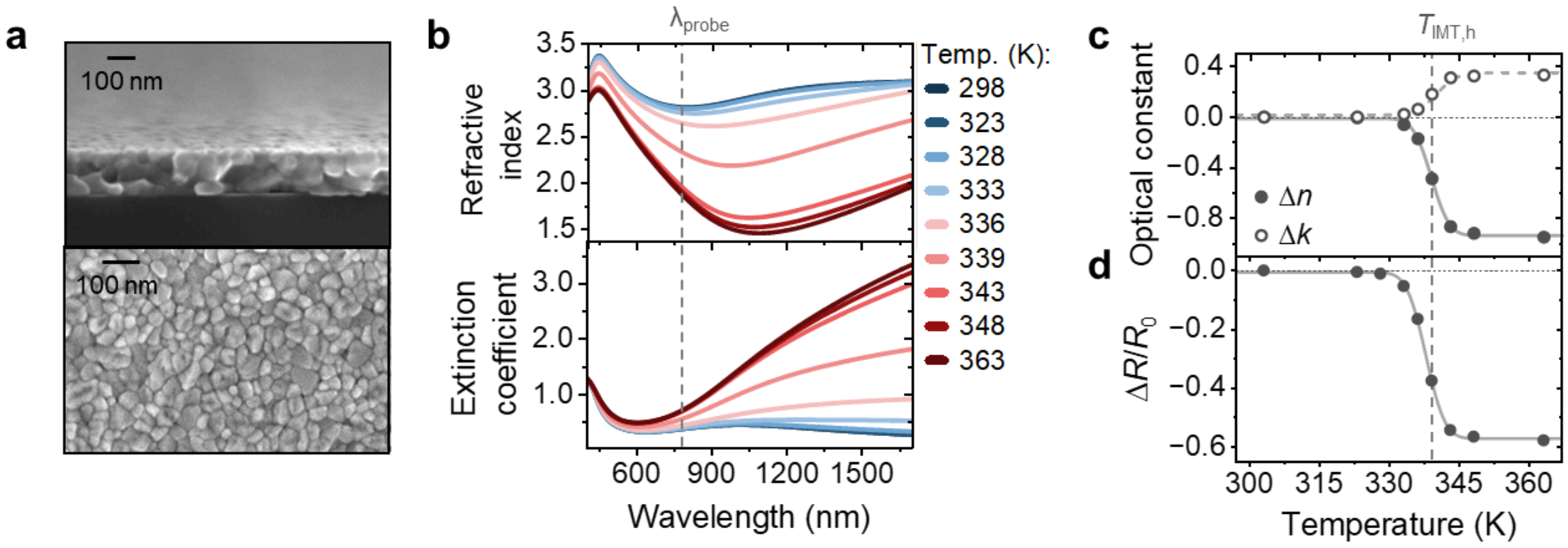

**Fig. 1. Morphological and optical characterization of $VO_2$ thin film.** (a) SEM side view and top view of $VO_2$ thin films. (b) Refractive index ($n$, top panel) and extinction coefficient ($k$, bottom panel) of a $VO_2$ thin film for increasing temperature. The vertical dashed gray line represents the probe wavelength. (c) Variation of the optical constants at 775 nm relative to 298 K, fit to Eq. (S1). (d) Reflectance change at 775 nm relative to 298 K simulated from optical constants in (c). The vertical dashed gray line represents the experimental phase transition temperature $T_{\mathrm{IMT,h}}$.

## Optical pump–probe imaging of phase domain

To trigger the $VO_2$ phase transition and spatiotemporally track the in-plane displacement of the metal–insulator boundary (Fig. 2a), we employed stroboscopic optical scattering microscopy (stroboSCAT) as developed by Delor *et al*.[37] This method has been used to visualize the propagation of charge carriers, heat, exciton-polaritons, phonon-polaritons, and waveguide modes at the nanoscale.[37–43] Here, we find that it is also naturally suited to imaging photoinduced phase domains in solid–solid phase-change materials, enabling access to mesoscale phase boundary transport emerging from the underlying polycrystalline microstructure.

In this experiment (Fig. 2a, see Methods), a 635 nm-wavelength light pulse of about 100 ps with a diffraction-limited spatial profile (full width at half maximum ≈ 350 nm) induces the insulator-to-metal phase transition in the $VO_2$ film. Electron–phonon coupling converts the deposited energy into heat faster than the duration of the hundreds-of-picosecond instrument response function.[23,44] At sufficiently high pump fluence, the resulting temperature jump locally exceeds $T_{\mathrm{IMT,h}}$. The isotherm at the radial position corresponding to the transition temperature represents the mesoscale interface between the metallic domain and surrounding insulating phase. We track the evolution of the transient metallic domain in space and time by imaging the resulting transient scattering profile in a reflection geometry ($\Delta R/R_0$) on a CMOS camera using a wide-field 775 nm-wavelength probe light pulse. The phase transition causes a strong change in the refractive index (Fig. 1c), providing a highly sensitive contrast mechanism with $\Delta R/R_0 \sim 10^{-1}$ (Fig. 2b)—orders of magnitude larger than the corresponding signal under electronic and thermal excitation in typical semiconductors or metals.[45–47] The photoinduced metallic domain surrounded by the insulating phase appears as a central principal feature—with negative or positive contrast depending on the $VO_2$ thickness and substrate—and a radially oscillating interference pattern (Fig. 2b, S4). The origin of this shape will be discussed later. Note that this approach is sensitive to the mesoscale phase boundary envelope, while the spatial resolution set by the point spread function of the probe averages over the nanoscale grain-level heterogeneity within the domain. The spatiotemporal evolution of this profile reports on the relaxation dynamics of the phase change.

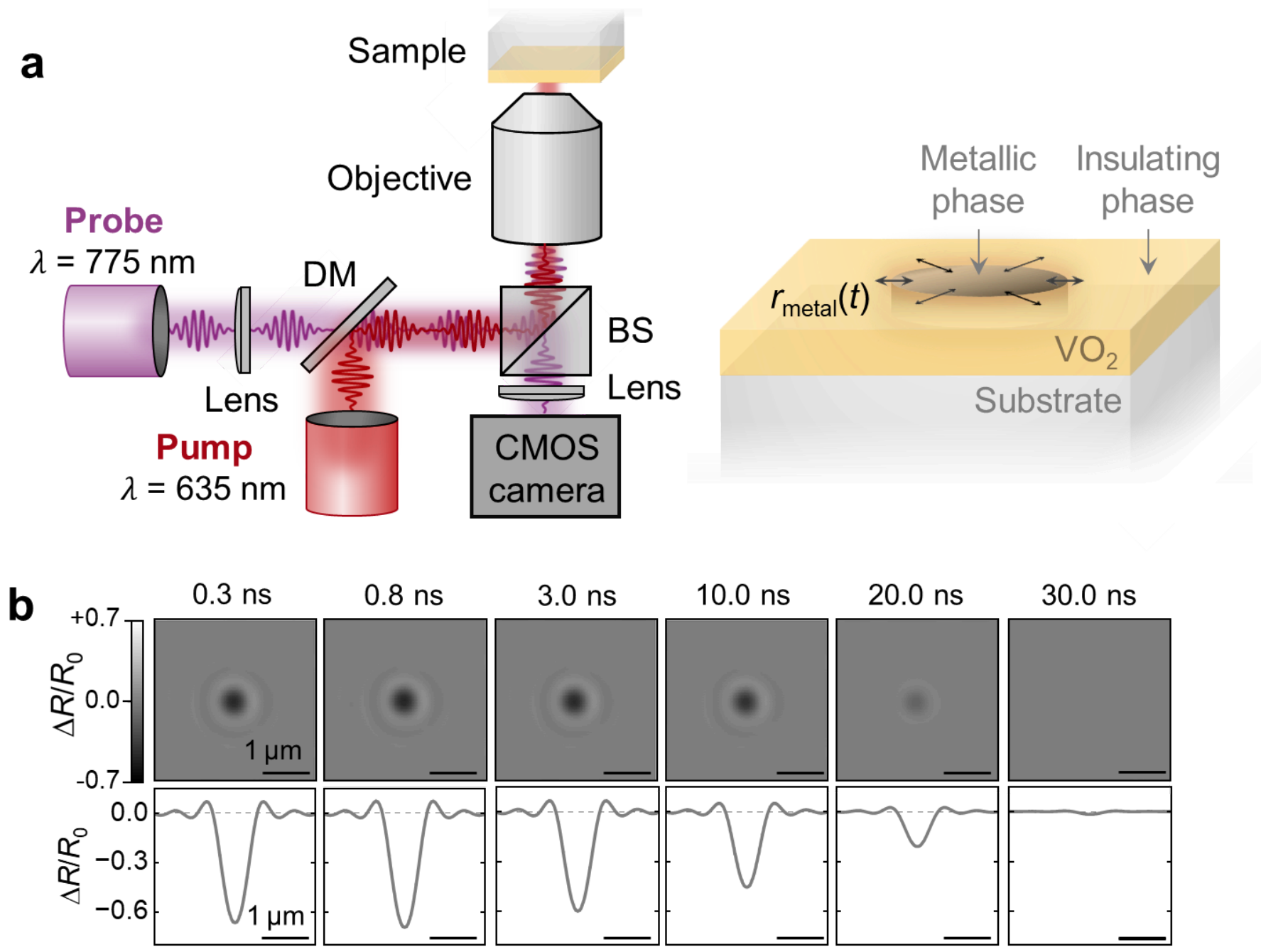

**Fig. 2. stroboSCAT technique.** (a) Schematic representation of the stroboSCAT experimental setup and observation of metallic phase domain propagation. DM labels the dichroic mirror and BS labels the 50/50 beam splitter. (b) Representative stroboSCAT data for selected pump–probe time delays after photoexcitation at 635 nm: top row shows spatially resolved maps of $\Delta R/R_0$ probed at 775 nm; bottom row displays the corresponding radial profiles, mirrored across the origin. Scale bars: 1 μm.

**Phase recovery kinetics limited by thermal dissipation**

Before examining the spatial evolution of the metallic phase, we focus on the relaxation kinetics (Fig. 3), which provides essential insights into the mechanism governing the phase transition dynamics. The rise of the signal is consistent with a 200 ps instrument response function (Fig. S5). The maximum reflectance change reaches a value in agreement with that predicted from the temperature-dependent optical constants (greater than 2-sigma of the hysteresis curve shown in Fig. 1d, S3) when the film is excited with a sufficiently high fluence of 7.3 mJ/cm$^2$ (evaluated using $1/e^2$ width of the pump spot—the fluence at the beam center is twice this value), while for lower excitation densities the phase transition may be only partially complete (optical constant change below 2-sigma of the hysteresis curve). Note that the sign and magnitude of the reflectance change are sensitive to the film thickness and substrate optical properties (Fig. 3, S4). Additionally, broadening due to the point spread function can also lead to an apparent reduction in $\Delta R/R_0$ for a phase domain with size close to the probe diffraction limit, despite local saturation in the optical constants at intermediate fluences (see Supplementary Section 3).

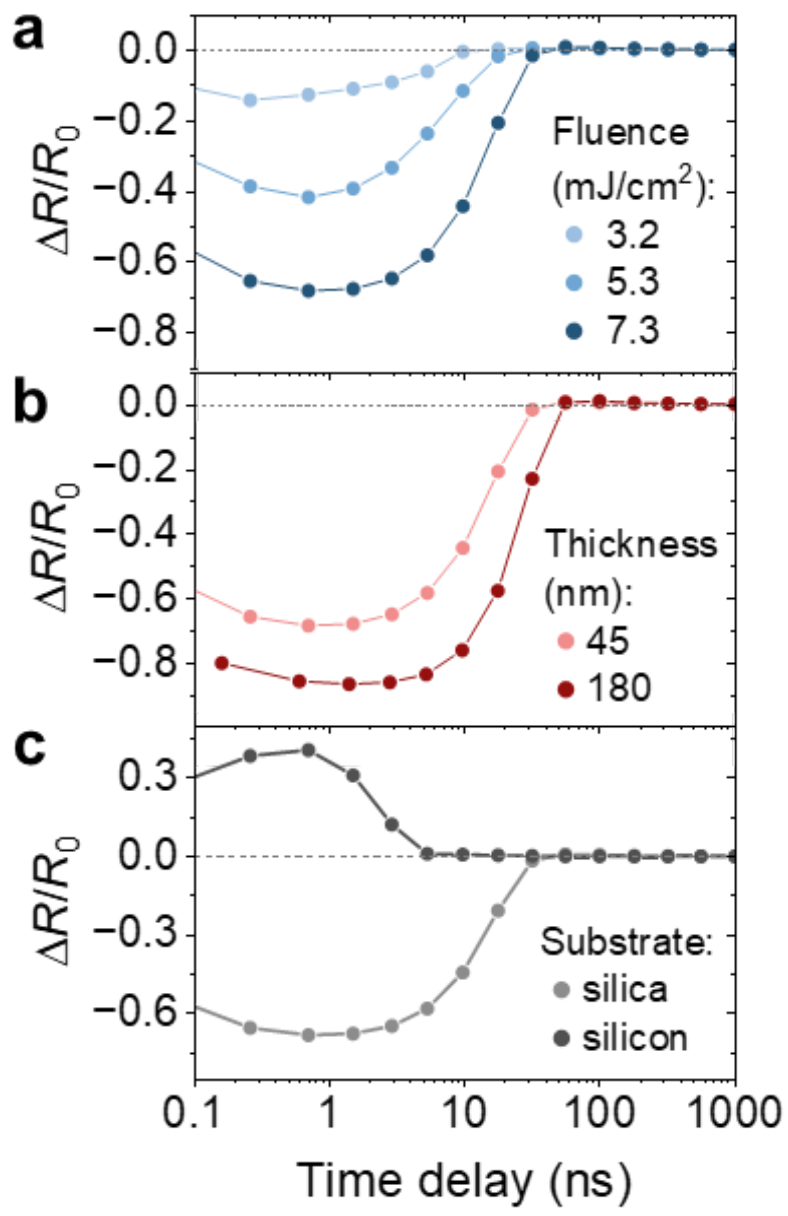

**Fig. 3. Influence of pump fluence, film thickness, and substrate on $VO_2$ phase relaxation kinetics.** Peak reflectance change time traces at 775 nm for (a) a 45 nm-thick $VO_2$ film on silica under three different excitation fluences, (b) $VO_2$ films on silica with thicknesses of 45 nm and 180 nm excited at a $1/e^2$ pump fluence of 7.3 mJ/cm$^2$, and (c) 45 nm-thick $VO_2$ films on silica and silicon substrates, excited at a pump fluence of 7.3 mJ/cm$^2$. The pump–probe time delay of 0 ns, not shown on the logarithmic time axis, was defined as the peak of the Gaussian function assumed for the instrument response function (IRF).

The decay of the peak reflectance variation is influenced by several experimental parameters (Fig. 3). First, pump fluence-dependent measurements reveal the relaxation time constant scales with the excitation energy density for both the 45 and 180 nm-thick films (Fig. 3a). Second, the time scale of the insulating phase recovery increases with film thickness for the same pump fluence (Fig. 3b). Third, we prepared identical $VO_2$ films with the same thickness (45 nm) on silica glass and silicon substrates. The use of a higher-thermal conductivity silicon substrate leads to a phase recovery that is substantially faster than that for a silica glass substrate (Fig. 3c). The half-life of each relaxation curve appears in Table S1, while additional data sets and alternate plots to compare lifetimes appear in Fig. S5. We note that control simulations confirmed that direct photoexcitation of the silicon substrate by residual pump energy transmitted through the $VO_2$ layer has negligible effect on both the pump–probe signal amplitude and the metallic domain radius evolution.

The above observations indicate that the relaxation of the metallic domain and recovery of the insulating phase are controlled by thermal dissipation to the substrate.[19,34] In this framework, a higher pump fluence results in greater heat injection into the film and therefore a longer time for heat dissipation to bring the film temperature below the phase transition threshold. Similarly, the thicker the film, the more heat must be transferred to the substrate to cool below the transition. Finally, this interpretation is further supported by the fact that using a substrate with higher thermal conductivity accelerates recovery. This phase relaxation regime has crucial consequences for the spatiotemporal dynamics of the metal–insulator phase boundary, as explored below.

**Spatiotemporal evolution of the metal-insulator boundary**

The stroboSCAT data for the $VO_2$ phase change deviate significantly from a simple time-broadened Gaussian description that is typically used for quasiparticle transport.[36,41,42] To extract the spatiotemporal evolution of the $VO_2$ metal–insulator boundary, an optical model that treats probe wave propagation in the presence of a non-Gaussian profile is essential. We implement a first-principles optical scattering model that calculates the $\Delta R/R_0$ image resulting from local changes in the sample complex refractive index $\Delta\tilde{n} = \Delta n + i\Delta k$ with arbitrary spatial profile (see Supplementary Information Section 3).[48] Although absorbed energy density originates from a Gaussian beam profile, the non-linear temperature dependence of the refractive index across the phase transition produces an in-plane refractive index modification that instead resembles a top-hat profile with transition-broadened edges (Fig. S7). This deviation from a Gaussian profile is especially apparent for peak refractive index changes greater than about 2-sigma along the temperature-induced hysteresis curve, inducing the saturation of the change near the center. Figures S7-S9 show this shape calculated from our experimental parameters, and in Fig. S10 we confirm such a profile for the reflectance profile under high fluence steady-state illumination.[49] We define an empirical analytical function for the refractive index profile that captures a smooth transition between the top-hat-like profile at saturation and a Gaussian-like profile as the domain crosses the phase transition (Eq. (1) in Methods section). We refer to the half-width at half the saturation value of the refractive index change profile ($\tilde{n}$ at $T_{\mathrm{IMT}}$ as measured by ellipsometry) as the effective metallic domain radius in what follows. We note that this definition represents a mesoscale picture of the multi-grain metallic domain based on the isotherm at the transition temperature.

Implemented in a numerical fitting procedure (see Methods), the model yields an accurate reconstruction of the experimental transient optical scattering profiles, including radial oscillations of the interference pattern (Fig. 4a,b and S11-S14). Visualizations of the underlying refractive index profiles output from the fits appear in Fig. 4a,b. We note that the fitted optical constants (Fig. S15) reach values consistent with the thermally-driven metallic phase seen in the ellipsometry measurements (Fig. 1b). Following this, the recovery of the refractive index to that of the insulating phase begins, following kinetics similar to the peak (or spatially integrated) reflectance change (Fig. 3), as shown in Fig. S15.

Importantly, we obtain the temporal evolution of the metallic domain size for different pump fluences in the different samples (Fig. 4c,d and S15). The high experimental signal-to-noise ratio, thanks to the extremely large optical contrast, allows few-nanometer precision in localizing the profile width. The spatiotemporal evolution of the metallic phase exhibits two stages: the domain size changes relatively slowly during the first several nanoseconds and then decreases abruptly until it vanishes (Fig. 4c,d, S15). The magnitude of the initial spatial displacement is up to tens of nanometers in the present experiments. The late-stage contraction exhibits an accelerating size decrease up until the metallic domain disappears. The ultimate timescale of disappearance increases with fluence. This overall behavior is in stark contrast to typical spatiotemporal measurements of electronic or thermal diffusion in semiconductors or metals, which instead exhibit monotonic outward motion and concave-up decay curves[41,42] unless population transfer occurs.[50,51] We interpret the above behavior as metal–insulator boundary propagation driven by heat flow, as explored further below.

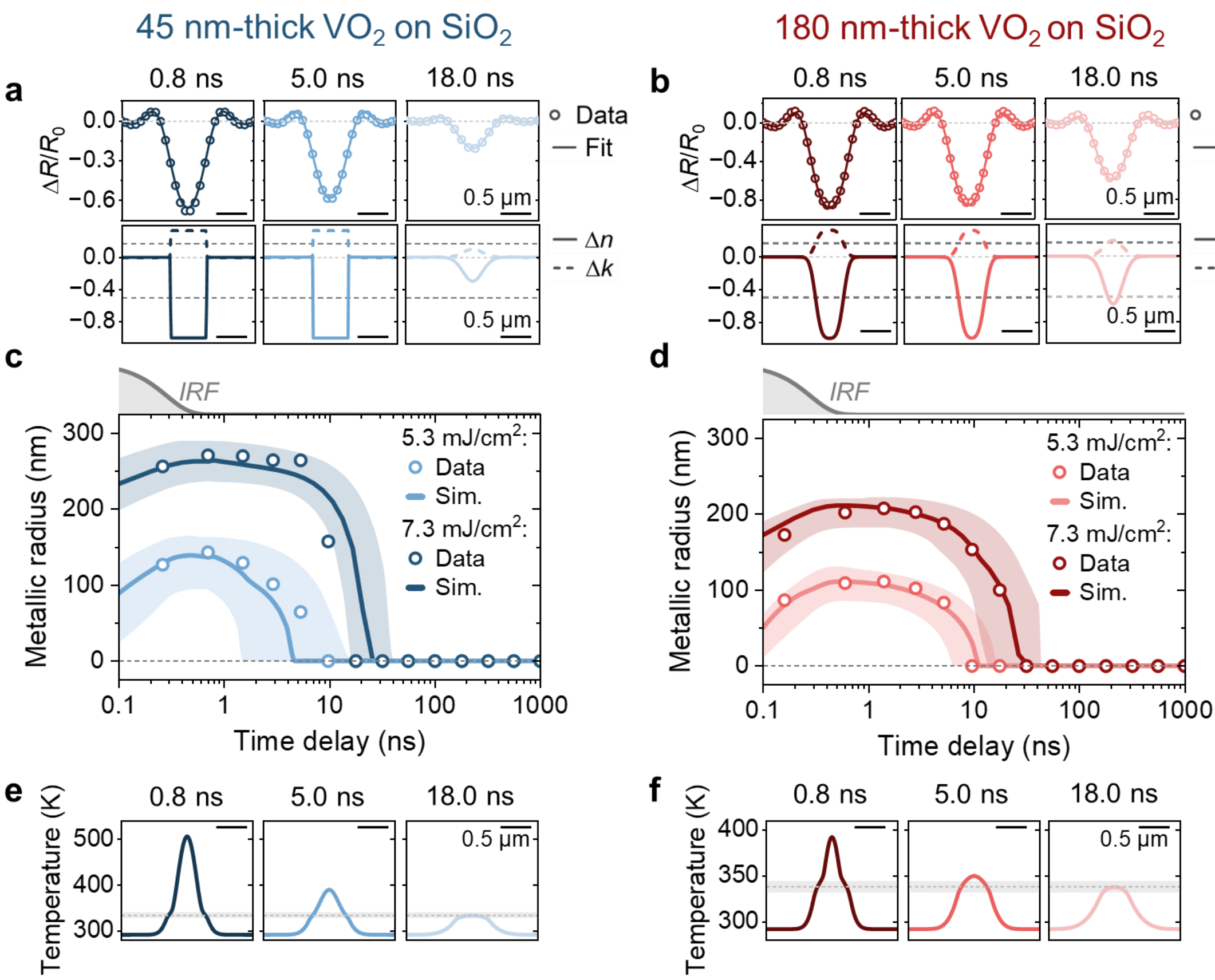

**Fig. 4. Fitting of the experimental data and simulations of metallic domain size evolution.** (a-b) Top row, experimental transient scattering profiles (empty dots) and fitting curves (solid lines) for selected time delays; Bottom row, the fitted change in the extinction coefficient (solid curves) and in the refractive index (dashed curves) relative to panel (a,b). The experimental data and the fits are reported for (a) 45 nm-thick and (b) 180 nm-thick $VO_2$ film on silica substrate excited at 7.3 mJ/cm$^2$. The gray dashed lines indicate where the changes in the refractive index and extinction coefficient exceed half of their saturation values, corresponding to the onset of the insulator-to-metal transition. (c-d) Experimental (empty dots) and simulated (solid lines) evolution of the metallic radius for (c) 45 nm-thick and (d) 180 nm-thick $VO_2$ film on silica substrate for two excitation fluences above the phase transition threshold. Experimental error bars on the radius (not visible) represent the standard deviation from the nonlinear least-squares fits to the optical scattering model. The shaded areas on simulation curves correspond to the evolution when evaluating the metallic radius considering transition broadening $\sigma$ = 8 and 4 K for 45 nm- and 180 nm-thick samples, respectively (Fig. S2). A 200 ps Gaussian instrument response function (IRF) used for simulations appears at the top of the plot for reference. (e-f) Simulated photoinduced temperature profile of the (e) 45 nm-thick and (f) 180 nm-thick $VO_2$ samples excited at 7.3 mJ/cm$^2$ relative to the simulations in panel (c,d) for selected time delays. The dashed grey line the indicates cooling transition temperature $T_{\mathrm{IMT,c}} \pm \sigma$. Scale bars in all subpanels represent 0.5 µm.

## Nonlinear thermal transport simulations

The spatiotemporal measurements provide key information to model the mechanism governing the phase boundary propagation under the present conditions. The overall phase relaxation behavior (Fig. 3) points to temperature-controlled dynamics. At the same time, the transport is nonlinear and nonmonotonic. To account for this behavior, we perform finite element simulations of the temperature profile evolution within the $VO_2$ layer taking into account both the latent heat of the phase transition and heat dissipation to the substrate. The initial temperature profile within the film is calculated based on the incident pulse energy of the Gaussian pump beam (350 nm full width at half maximum), the optical constants, the heating transition temperature $T_{IMT,h}$, and specific heat capacity curve accounting for latent heat, assuming a uniform profile in depth. This temperature profile is allowed to evolve by Fourier's law subject to the sample and substrate geometry and temperature-dependent thermal properties. The model assumes diffusive thermal transport based on the strongly scattering nature expected for the polycrystalline thin films with grain size well below the resolution of the measurement. As experimental inputs, the simulation considers the cooling transition temperature $T_{IMT,c}$ obtained from transmittance hysteresis measurements (330 K and 334 K for the 45 nm- and 180 nm-thick films, respectively, Fig. S2), as well as the thermal conductivities of $VO_2$ in both phases as determined by modulated thermoreflectance (Fig. S16-S18, Table S2). The values are consistent with a polycrystalline film.[52] The latter technique also yields the interfacial resistance between $VO_2$ and the substrate, which influences heat dissipation. We note that the lattice and electronic contributions to the thermal conductivity are implicitly accounted for by using the experimental thermal conductivities. Additional parameters, including the density and latent heat associated with the insulator–metal phase transition, were taken from literature.[33,53] In particular, the specific heat capacity curve exhibits a peak centered at the transition temperature, with an integrated area corresponding to the latent heat (Fig. S18).[54] This causes a sharp decrease in the thermal diffusivity around the temperature of the phase transition (Fig. S18), which modulates the expansion of the phase boundary. It is this strongly temperature-dependent thermal diffusivity that induces the nonlinear thermal transport despite a local Fourier's law description.

The initial simulated in-plane temperature profile exhibits a shoulder at the temperature corresponding to the phase transition due to the latent heat involved in photoinduced heating (Fig. 4e,f, S19-S21). This non-Gaussian shoulder becomes more accentuated throughout the subsequent evolution as the latent heat at that point slows thermal transport and delays the local temperature changes. We assume that the system is in a regime where the phase front closely follows the isotherm $T = T_{IMT}$;[55] thus, to obtain the simulated metallic domain size, $r_{metal}(t)$, we take the average radial distance at which the temperature equals the transition threshold (gray dashed line in Fig. 4e,f). This definition for the simulated radius corresponds to the experimental radius that is in turn defined by the inflection point of the sharp, nonlinear refractive index change at the same temperature.

The magnitude and kinetics of the simulated metallic domain size are in good agreement with the experimental data for all samples (Fig. 4c,d, S19-S22). We emphasize that such agreement was achieved without parameter fitting, but rather solely relying on experimental or literature values. This suggests that the continuum model simulations capture the essential mesoscale physics of the problem, wherein a quasi-equilibrium coexistence of phases is mediated by local heat diffusion.

Motivated by the agreement with the data, we use the simulations as a platform to explore the nature of phase boundary propagation driven by the nonlinear evolution of the thermal gradient. To illustrate the physical picture guiding the following analysis, Fig. 5a presents a schematic depiction of the model. To isolate the transport behavior that is inherent to heat flow in $VO_2$, we consider an isolated film—to eliminate effects from heat dissipation to the substrate—and a uniform profile across the film thickness. We simulate the metallic domain size evolution for a range of peak temperatures. We find that the initial slope of radius versus time curves can be either positive (domain growth) or negative (domain reduction), depending on the initial temperature rise (Fig. 5b).

This occurs because, when the peak temperature significantly exceeds $T_{IMT}$, the transition is located at the wings of the profile such the lateral initial spreading of the temperature profile pushes the transition isotherm outward faster than the dissipation can cause the temperature to lower, resulting in a transient outward propagation of the phase boundary. In contrast, at lower peak temperatures, the transition is located near the center of the profile where lateral heat spreading only causes the local temperature to decrease, so the domain shrinks monotonically from the outset. This balance is captured analytically in Supplemental Section 6 (Fig. S23) for the simplified case of temperature-independent diffusivity, where the condition for initial outward propagation is shown to correspond to $\Delta T_{peak} > e\Delta T_{IMT}$ (differences relative to ambient temperature).

The transport regime is revealed by the time dependence of the phase boundary position. The initial transport is approximately diffusive, showing a nearly linear time dependence of the squared position ($r^2_{metal}(t) - r^2_{metal}(t_0) = 2D_{eff}t$) before reversing direction. The final transport time dependence leading into disappearance is also consistent with diffusion. This motivates an examination of the instantaneous effective diffusivity curves extracted from the simulations, defined as $D_{eff}(t) = \mathrm{d}/\mathrm{d}t(r^2_{metal}(t))/2$, appearing in Fig. 5c and S24 (with sign encoding the direction of transport). Interestingly, the phase boundary, despite being locally associated with the transition temperature and corresponding thermal properties, can move faster than the latent heat-modulated thermal diffusivity at the transition temperature ($D_L(T_{IMT}) \approx 155$ nm$^2$ ns$^{-1}$, Fig. 5c, S18). Expansion can moreover proceed faster than the thermal diffusivities of both the pure insulating and metallic phases ($D \approx 800$–$1000$ nm$^2$ ns$^{-1}$, Fig. 5c, S18). We note that while the time derivative analysis makes experimental instantaneous diffusivity prone to error, the result supports the conclusions of the simulations (Fig. S25).

The final reduction of the metallic domain is due to the temperature profile spreading outward in the $VO_2$ layer, causing the radial position corresponding to the temperature threshold of the phase transition to move progressively toward the center of the spot (Fig. 5a). Interestingly, the final shrinking behavior, reached some time before the metallic domain disappears, converges to a common terminal diffusivity that is independent of the initial temperature rise (Fig. 5c, S24). Surprisingly, its magnitude also exceeds $D_L(T_{IMT})$. Specifically, latent heat appears to bound the domain reduction rate to twice the latent-heat modulated diffusivity, $2D_L(T_{IMT})$ (Fig. 5c).

The observation that phase-boundary motion can proceed substantially faster than would be expected from thermal diffusion, despite being driven by the evolution of the thermal gradient itself, is a key result of this work. This phenomenon occurs because, while the thermal diffusivity governs the width of the temperature profile, the quasi-equilibrium phase boundary corresponds to a fixed temperature isotherm $T_{IMT}$ along that profile (Fig. 5a). The radial motion of a given amplitude along a temperature gradient (here $r(t)$ at $T = T_{IMT}$) is a fundamentally different quantity than the profile width. As such, its magnitude is not restricted to the diffusivity and can be significantly larger. We provide a simple analytical illustration of this phenomenon for a diffusing Gaussian profile without latent heat in Supplemental Material Section 6 (Fig. S23), while the full nonlinear simulations (Fig. 5b) demonstrate that it survives in the presence of the latent heat peak.

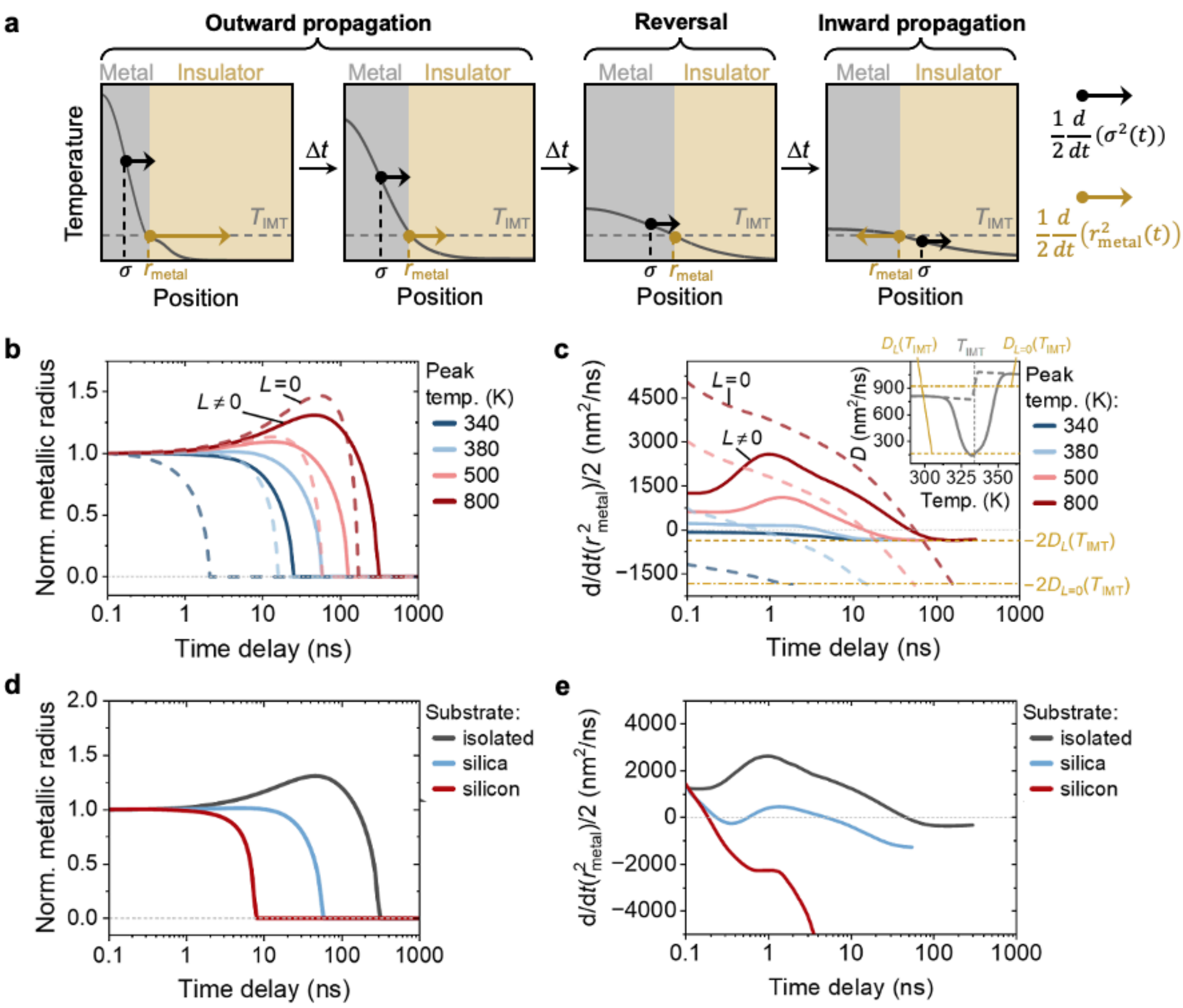

**Fig. 5. Simulations of metallic domain size evolution.** (a) Schematic depiction of temperature profile and phase boundary evolution including stages of propagation outward, reversal, and inward. The metallic and insulating domains are denoted by grey and yellow shading, respectively. The radial positions of the phase domain boundary, $r_{\text{metal}}$, and the standard deviation of the temperature profile, σ, are indicated along the $x$-axis. The corresponding direction of propagation and magnitude of the instantaneous diffusivities of the overall temperature profile, $d/dt(\sigma^2(t))/2$, and the phase domain boundary, $D_{\text{eff}}(t) = d/dt(r^2_{metal}(t))/2$, are represented by the direction and length of the black and yellow arrows, respectively. (b) Simulated metallic radius normalized to the initial value for different peak temperatures, taking a uniform out-of-plane profile and no substrate. The solid and dashed curves show the evolution with and without the latent heat, $L$, respectively. (c) Instantaneous effective diffusivity curves $D_{\text{eff}}(t)$ with and without latent heat extracted from the simulations in panel (b), defined as $D_{\text{eff}}(t) = d/dt(r^2_{metal}(t))/2$ (with sign encoding direction) based on the mean-squared displacement for a diffusive process. Inset: Input diffusivity versus temperature in the vicinity of the phase transition with and without the latent heat peak in heat capacity included. The magnitudes of the diffusivities at the phase transition temperature ($T_{\text{IMT}}$) with ($D_L(T_{\text{IMT}}) = 155$ nm$^2$/ns) and without ($D_{L=0}(T_{\text{IMT}}) = 940$ nm$^2$/ns) latent heat are indicated for reference. (d,e) Influence of substrate on (d) metallic radius evolution and (e) instantaneous effective diffusivity for $VO_2$ layers that are isolated from the environment or in contact with silica or silicon substrates. Latent heat and interfacial thermal resistance ($1.0 \times 10^{-8}$ m$^2$ K W$^{-1}$) are accounted for in these simulations, with a peak temperature rise of 800 K.

Heat transfer to the substrate leads to a shorter-lived metallic domain (Fig. 5d) and accelerates the instantaneous rate of domain collapse (Fig. 5e). The inward transport can thus greatly exceed the thermal diffusivity of $VO_2$ despite the inertial effect of latent heat, with a magnitude that increases for increasing substrate thermal conductivity.

The finite element simulations further provide a simple way to evaluate the role of latent heat in the transport. Fig. 5b shows the simulated metallic domain size evolution with and without the latent heat included in the heat capacity curve. The comparison shows how the latent heat acts as thermal inertia, both slowing down phase front propagation (both outward and inward) and delaying the disappearance of the metallic domain. In both cases, the initial propagation exhibits a logarithmic dependence on initial temperature with potential to exceed the pure phase thermal diffusivity. In contrast to the constant diffusivity late-stage collapse observed in the presence of latent heat, our simplified analytical model (Supplemental Material Section 6, Fig. S23) and simulations (Fig. 5c) indicate that, in the absence of latent heat and substrate, the inward transport accelerates progressively until disappearance. However, time-dependent diffusive transport remains the best description of the regime, as the collapse reaches precisely twice the average diffusivity of the pure phases around the transition, $2D_{L=0}(T_{\mathrm{IMT}})$, independent of the initial temperature rise. This scenario highlights the distinct behaviors expected for first and second order phase transitions.

We note that while one may be tempted to examine the instantaneous velocity, as growth rate has been invoked in the past,[17,18,29] we find that this quantity varies continuously throughout the entire dynamics (Fig. S24). The instantaneous diffusivity, in contrast, exhibits a temporarily stable magnitude both at the beginning and end of the domain growth and reduction when latent heat is involved, and otherwise has a consistent value just before disappearance for phase transitions without latent heat (Fig. 5c). The final diffusivity is moreover independent of the initial temperature. Thus, under the condition of thermal gradient-driven phase boundary propagation, phase front diffusivity emerges as a more fundamental quantity than velocity, the latter being simply phenomenological.

**Discussion**

We have presented an observation of and mechanistic picture for nonlinear thermally-driven insulator–metal phase-boundary motion in $VO_2$. The use of a diffraction-limited photoinduced heat pulse allowed for optical imaging of both the initial outward propagation and late-time contraction of a transient metallic domain—as well as its ultimate disappearance—with sub-micron spatial and sub-nanosecond temporal resolution. The agreement between experiments and parameter-free nonlinear thermal transport simulations identifies the phase boundary propagation and metal-to-insulator relaxation as being driven by the evolution of the thermal gradient. In this regime, the observed domain kinetics arise from the nonlinear propagation of heat through the film, modulated by the latent heat of the first-order transition and heat transfer across the film–substrate interface. As such, the instantaneous diffusivity of phase boundary propagation is a dynamic quantity that can take on a range of magnitudes and encompass both growth and reduction throughout the course of relaxation. Latent heat acts to suppress transport and prolong the coexistence of phases. Yet, the phase front is able to propagate significantly faster than the thermal diffusivity at the transition temperature, despite being carried by the thermal gradient itself. This latter phenomenon should be general to thermal gradient-driven phase transitions. Taken together, these results establish a physical framework connecting latent heat, nonlinear heat transport, and mesoscale phase-front kinetics.

Our results also establish a link between heat dissipation and mesoscale phase-front motion. This implies that controlling interfacial heat extraction, film thickness, and excitation geometry offers a direct route to tailoring the temporal and spatial evolution of phase transitions. Taken together, the observed phenomena define a framework for engineering the spatiotemporal response essential to the design of phase-change devices (see Supplemental Material Section 9 for a discussion on further implications). For example, we anticipate that the strong refractive index change of the phase change, combined with control over phase-front propagation, will open opportunities for on-demand transient optical patterning and the design of photoinduced optical metamaterials.

Extending beyond carrier diffusion and thermoreflectance measurements, our results demonstrate optical pump–probe microscopy as a powerful method to spatiotemporally resolve phase boundary motion with strong contrast—a high spatiotemporal resolution analogy to Burfoot and Parker furnace measurements.[55,56] We emphasize that the use of an optical scattering model was essential to describe the shape of the pump–probe images due the strong changes in both the real and imaginary parts of the refractive index as well as a diffraction-limited and non-Gaussian underlying profile.

Beyond $VO_2$, we expect the tools and findings to be widely applicable to other first order as well as second order solid–solid phase transitions. This could be a platform for examining analytical and phase-field treatments of moving boundary problems in solid–solid transitions. Moreover, extending to thermally-driven phenomena beyond phase-change materials, the spatiotemporal scales accessible here may be compatible with measuring non-diffusive heat transfer[42,57] or collective thermal transport phenomena such as second sound.[58]

## Methods

### Sample synthesis and characterization

Polycrystalline $VO_2$ thin films were deposited on silica or silicon substrates by reactive magnetron sputtering from a vanadium target in an $Ar/O_2$ atmosphere. The as-deposited amorphous films underwent a post-deposition annealing at 450 °C in a $N_2$ atmosphere to improve the crystalline quality, yielding the characteristic $VO_2$ monoclinic phase at room temperature.

The thickness of the $VO_2$ polycrystalline thin films was characterized using a KLA Tencor P-17 profilometer, while the morphology was studied with a Zeiss Sigma Gemini HD field-emission scanning electron microscope and a NT-MDT Solver Pro atomic force microscope operating in semi-contact mode. The structural characterization was performed with a Panalytical X'Pert Pro X-ray diffractometer equipped with an Anton Paar DHS900 sample holder to perform temperature-dependent measurements. The transmittance spectra were acquired with a JASCO V-670 UV-Vis-NIR spectrophotometer, while the optical constants were measured using a J.A. Wollam VASE ellipsometer. Both these instruments were equipped with heatable sample holders to perform measurements below and above the transition temperature. No significant differences were evidenced in the films grown on the two investigated substrates. Detailed characterization results are reported in the Supplemental Material Section 1.

### Finite element simulations

*Reflectance spectrum starting from ellipsometry*: The optical response (reflectance and absorbance) of $VO_2$ thin films with varying thicknesses was simulated using the Wave Optics Module of COMSOL Multiphysics 6.2. A two-dimensional geometry with periodic boundary conditions was employed to optimize computational efficiency. The optical constants of the $VO_2$ layer were those determined experimentally by ellipsometry, while the substrate properties ($SiO_2$ or Si) were defined using the built-in dielectric data. The incident light was modeled as a plane wave launched through a port node. The total reflected spectra at normal incidence were obtained by collecting the backpropagation electric field that enters the same port. The reflection of the back side of the substrate is neglected, while Perfectly Matched Layers (PMLs) were implemented for both the interfaces with the substrate and the air (or oil) to ensure accurate treatment of outgoing radiation.

*Temperature profile evolution*: The finite element simulations of nonlinear heat transport were performed with the commercial software COMSOL Multiphysics 6.2 using the Heat Transfer in Solids module. A two-dimensional geometry with an axisymmetric model helps to significantly cut down the time required for calculations. A thin $VO_2$ layer was placed on top of either a silica or silicon substrate. Input values for the phase-dependent (temperature-dependent) thermal conductivities, phase transition temperatures, and the interfacial thermal resistance were obtained from dedicated experiments (Fig. S2, S16, S17, Table S2), while other properties (density, heat capacity, and latent heat) were taken from literature (Fig. S18).[33] The thermal conductivity and heat capacity account for sigmoidal change around the phase transition temperature, and the latent heat is included via a peak in heat capacity (Fig. S18). The upper boundary of the film was modeled as thermally insulating to account for the negligible heat exchange with the low thermal conductivity immersion oil. At the film–substrate interface, thermal coupling was represented by an equivalent resistive layer with the interfacial resistance taken from modulated thermoreflectance measurements (Supplementary Section 4). The mesh was constructed using an "extremely fine" element size. The finite element simulation solves the heat equation (Fourier's law) using a quadratic Lagrange discretization to calculate the time dependence of the temperature profile with a relative tolerance of $10^{-7}$.

**stroboSCAT measurements**

The experimental setup for stroboscopic optical scattering microscopy (stroboSCAT) was previously described in detail[59] and was based on an instrument developed by Delor *et al.*[37] Briefly, the home-built microscope used an inverted microscope configuration with a 1.4 numerical aperture oil-immersion objective. Two separate laser diodes were used for pump and probe sources, operated by drivers from PicoQuant to control the pump–probe time delay. Here we used laser heads with center wavelengths of 635 nm and 775 nm (LDH-D-C-640, LDH-D-C-780, PicoQuant) for the pump and probe, respectively. The rise of the signal (Fig. S5) was fit to a Gaussian instrument response function (IRF), giving a pump–probe IRF standard deviation of 200 ps, suggesting average pump and probe pulse durations of 140 ps, consistent with the nominal pulse duration of ≈100 ps. We used a laser repetition rate of 500 kHz, with the pump modulated at 660 Hz. Both the pump and probe were spatially filtered through pinholes (75 µm and 100 µm, respectively), the probe beam passed through a lens to focus on the back focal plane of the objective, the two beams are overlapped using a dichroic mirror and reflected into the objective by a 50/50 beamsplitter. This leads to an overlapped near-diffraction-limited pump (full width at half maximum ≈ 350 nm) and wide-field probe (≈10 µm) illumination. Probe light reflected from the sample was collected through the same objective, isolated with a band-pass filter and focused onto a CMOS camera, triggered at 660 Hz by the laser driver. The pixel calibration was done using a 1951 USAF Pattern resolution target (R3L1S4PR, Thorlabs), giving 28 nm/pixel. stroboSCAT images are generated by taking the difference between pump-on and pump-off raw pixel intensities, normalized to the raw pump-off intensities, yielding $\Delta R/R_0$ contrast images as a function of time delay. Data acquisition was implemented in LabVIEW 2022 64-bit. Data analysis and plotting were performed using Origin and Python. Fitting to obtain the refractive index change and metallic domain size is described in the following section.

The pump fluence was found by measuring the power arriving at the sample—using a microscope slide format power sensor (S171C, Thorlabs) and power meter (PM100D, Thorlabs)—and the beam spot size by imaging its reflection. Experiments were conducted at an ambient temperature of 292 K.

**Optical scattering model**

The optical scattering model used to fit the pump-probe images was originally presented for transmission-based measurements[48] and was adapted to the present experimental setup with a reflection configuration (Fig. S6). Briefly, this model simulates how the reflectance image of a film is influenced by the photoinduced modifications of the complex refractive index, $\Delta\tilde{n} = \Delta n + i\Delta k$, relative to the background $\tilde{n}_0 = n_0 + ik_0$. The locally modified refractive index alters the phase and amplitude of the probe pulse as it propagates through the film. The interference between the perturbed portion of the probe and the background reflected probe leads to a spatial interference pattern in the intensity captured by the camera. This description is analogous to scattered and reference fields in interferometric scattering microscopy (iSCAT)[60,61], but the present optical model accounts for an arbitrary profile for $\Delta\tilde{n}$. The resulting pump–probe image displays a central peak accompanied by radial oscillations that arise from the interference, as seen in experiments. Below we present the derivation of the model.

The first step is to define a functional form for $\Delta\tilde{n}(r)$. As detailed in Supplementary Section 3 (Fig. S7-S10), the nonlinear nature of the $VO_2$ phase transition gives rise to a profile for the optical constant variations that closely resembles a top-hat distribution with broadened edges. Accordingly, the chosen expression for $\Delta\tilde{n}(r)$ is:

$$\Delta\tilde{n}(r;\Delta n,\Delta k,\delta,\varepsilon)=(\Delta n+i\Delta k)\frac{1}{2}\left[\mathrm{erf}\left(\frac{r+\delta}{\sqrt{2}\,\varepsilon}\right)-\mathrm{erf}\left(\frac{r-\delta}{\sqrt{2}\,\varepsilon}\right)\right] \tag{1}$$

where $\delta$ denotes the radius of the plateau and $\varepsilon$ corresponds to the broadening at the edges of the top-hat profile. The experimental observable—the refractive index change—is related to the portion of the film for which the phase transition occurs. The effective metallic domain radius corresponds to the position $r$ when $\Delta\tilde{n}(r)$ equals half of the saturation value shown in Fig. 1c. For $\varepsilon \ll \delta$, the metallic radius is conveniently $\approx\delta$, while for $\varepsilon \approx \delta$ the metallic domain radius is smaller than $\delta$ and must be found numerically. Note that $\varepsilon$ captures both the intrinsic broadening of the temperature-dependent refractive index profile (Fig. S2) and the (time-dependent) steepness of the temperature gradient in the vicinity of the transition temperature. Equation (1) assumes a uniform profile across the film thickness.

The reflected near-field electric field $E^{nf}(r)$ is obtained by multiplying the incident field with the Fresnel total reflection coefficient. This coefficient is evaluated using the transfer matrix method, which accounts for the propagation inside each layer and the different interfaces. In particular, the forward- and backward-propagating plane waves ($E^+$ and $E^-$, respectively) of the $m$-th layer are related to the ones of the ($m$+1)-th layer through the relation

$$\begin{pmatrix}E_m^+\\E_m^-\end{pmatrix}=M_R^{m,m+1}\begin{pmatrix}E_{m+1}^+\\E_{m+1}^-\end{pmatrix}=\begin{pmatrix}1 & r_{m,m+1}\\ r_{m,m+1} & 1\end{pmatrix}\frac{1}{t_{m,m+1}}\begin{pmatrix}E_{m+1}^+\\E_{m+1}^-\end{pmatrix}, \tag{2}$$

while the propagation inside the $m$-th layer is given by a matrix element of the form

$$M_P^m=\begin{pmatrix}e^{-i\gamma_m} & 0\\ 0 & e^{i\gamma_m}\end{pmatrix} \tag{3}$$

where $\gamma_m = 2\pi\tilde{n}_m L_m/\lambda$, and $L_m$ is the thickness of the $m$-th layer.

The matrix $M_P$ expresses the phase accumulation when the electric field propagates inside layer $m$, while matrix $M_R$ describes the contribution of each interface through the Fresnel reflection ($r_{m,m+1}$) and transmission ($t_{m,m+1}$) coefficients. As shown in the scheme reported in Fig. S6, three interfaces are considered (oil/$VO_2$, $VO_2$/substrate, and substrate/air), while both the oil and air layers are treated as semi-infinite media (only $M_R$ is present for these terms). Therefore, the previous relation becomes

$$\begin{pmatrix} E_{\text{oil}}^{+} \\ E_{\text{oil}}^{-} \end{pmatrix} = M_R^{\text{oil,VO}_2} \cdot M_P^{\text{VO}_2} \cdot M_R^{\text{VO}_2\text{,substrate}} \cdot M_P^{\text{substrate}} \cdot M_R^{\text{substrate,air}} \begin{pmatrix} E_{\text{air}}^{+} \\ E_{\text{air}}^{-} \end{pmatrix} =$$

$$= \begin{pmatrix} 1 & r_{\text{oil,VO}_2} \\ r_{\text{oil,VO}_2} & 1 \end{pmatrix} \frac{1}{t_{\text{oil,VO}_2}} \begin{pmatrix} e^{-i\gamma_{\text{VO}_2}} & 0 \\ 0 & e^{i\gamma_{\text{VO}_2}} \end{pmatrix} \cdot$$

$$\cdot \begin{pmatrix} 1 & r_{\text{VO}_2\text{,substrate}} \\ r_{\text{VO}_2\text{,substrate}} & 1 \end{pmatrix} \frac{1}{t_{\text{VO}_2\text{,substrate}}} \cdot$$

$$\cdot \begin{pmatrix} e^{-i\gamma_{\text{substrate}}} & 0 \\ 0 & e^{i\gamma_{\text{substrate}}} \end{pmatrix} \begin{pmatrix} 1 & r_{\text{substrate,air}} \\ r_{\text{substrate,air}} & 1 \end{pmatrix} \frac{1}{t_{\text{substrate,air}}} \begin{pmatrix} E_{\text{air}}^{+} \\ E_{\text{air}}^{-} \end{pmatrix} =$$

$$= M^{tot} \begin{pmatrix} E_{\text{air}}^{+} \\ E_{\text{air}}^{-} \end{pmatrix} \tag{4}$$

Since there is no backward wave returning from the air layer, $E_{air}^{-} = 0$, hence $E_{oil}^{+} = M_{1,1}^{tot} E_{air}^{+}$ and $E_{oil}^{-} = M_{2,1}^{tot} E_{air}^{+}$. These two quantities are then used to evaluate the overall reflection coefficient as

$$r_{tot} = \frac{E_{\text{oil}}^{-}}{E_{\text{oil}}^{+}} = \frac{M_{2,1}^{tot}}{M_{1,1}^{tot}}, \tag{5}$$

which directly corresponds to the near-field electric field $E^{nf}(r)$, given that the incident field is normalized to unity. To propagate the near-field electric field to the far field at the objective lens, its Fourier transform (Hankel transform in 2D) is computed: $E^{ff}\left(k_{ff}\right) = \boldsymbol{\mathcal{F}}\left(E^{nf}(r)\right)$. Additionally, a phase factor $e^{i\Delta z\sqrt{2\pi/\lambda - k_{ff}}}$ is included to study how the far-field electric field varies when different imaging planes within the object space are considered. In our setup, the defocusing $\Delta z$ is achieved by translating the sample along the $z$-axis using a piezoelectric stage. The higher spatial frequencies are filtered according to the numerical aperture ($NA$) of the objective. Specifically, only the electric field components $E_{filtered}^{ff}\left(k_{ff}\right)$ such that the spatial frequencies satisfies the condition $|k_{ff}| \leq \frac{2\pi}{\lambda}NA$ are transmitted. Finally, the field at the image plane is computed through an inverse Fourier transform $E^{im}(r) = \boldsymbol{\mathcal{F}}^{-1}\left(E_{filtered}^{ff}\left(k_{ff}\right)\right)$. This quantity is directly related to the relative change in reflectance:

$$\frac{\Delta R}{R_0}(r;\delta,\varepsilon,\Delta z,\Delta n,\Delta k) = \frac{\left|E^{im}(r;\delta,\varepsilon,z_0+\Delta z,n_0+\Delta n,k_0+\Delta k)\right|^2}{\left|E^{im}(r;\delta,\varepsilon,z_0,n_0,k_0)\right|^2} - 1 \tag{6}$$

This expression indicates that the modification of the optical response depends on the amplitude and spatial extent of the variations in the real and imaginary parts of the complex refractive index, as well as on the degree of defocusing. These quantities are treated as fitting parameters, enabling accurate reproduction of the experimental profiles and retrieval of their temporal evolution. The fitting procedure and results for some examples are reported in Supplementary Section 3.

## Acknowledgements

This work was supported by ERC grant InSituDynamics (grant no. 101218874) and the Emergence Research Programme of Alliance Sorbonne Université. This work was performed in the framework of the PNRM Project “METEORE” (contract n. 564, year 2021) funded by the Italian Ministry of Defense and was supported in part by the Italian MUR Departments of Excellence grant 2023−2027 “Quantum Frontiers”.

**SUPPLEMENTARY INFORMATION**

**for**

# Spatiotemporal Imaging of Thermally Driven Metal–Insulator Phase Front Propagation in $VO_2$

Ilaria Delbono[1,2], Raouf Amara[3,4], Matias Feldman[1], Hankun Yang[3], Cédric Leau[1], Carlo Scian[2], Erwan Dandeu[1], Sébastien Royer[1], Debora Pierucci[1], Emmanuel Lhuillier[1], Franck Vidal[1], Tiziana Cesca[2], Hilton B. de Aguiar[3], Jose Ordonez-Miranda[1], Danièle Fournier[1], Raj Pandya[4,5], Boris Kalinic[2]*, Giovanni Mattei[2]*, James K. Utterback[1]*

[1] Sorbonne Université, CNRS, Institut des NanoSciences de Paris, 75005 Paris, France
[2] Physics and Astronomy Department, University of Padova, via Marzolo 8, I-35131 Padova, Italy
[3] Laboratoire Kastler Brossel, ENS-Université PSL, CNRS, Sorbonne Université, Collège de France, Paris, France
[4] Department of Chemistry, University of Warwick, Coventry, UK
[5] Yusuf Hamied Department of Chemistry, University of Cambridge, Cambridge CB2 1EG, United Kingdom

*E-mail: boris.kalinic@unipd.it
*E-mail: giovanni.mattei@unipd.it
*E-mail: james.utterback@sorbonne-universite.fr

**Table of Contents**

## SUPPLEMENTARY SECTION 1:
## Characterization of $VO_2$ thin films

Characterization of the polycrystalline $VO_2$ films was performed using scanning electron microscopy (SEM), atomic force microscopy (AFM), and grazing-incidence X-ray diffraction (GI-XRD). The morphology was investigated by SEM and AFM, as shown in Fig. 1 and S1a, revealing grains with an average diameter of 50 nm and a root-mean-square surface roughness of 3 ± 1 nm. In Fig. 2b, GI-XRD was employed at 295 K and 365 K to assess the crystalline quality of the samples by identifying diffraction peaks corresponding to both the low-temperature insulating monoclinic phase and high-temperature metallic rutile phase, respectively. The inset of Fig. S1b highlights the shift of the (011) and (110) peaks, confirming the structural transition from the monoclinic to the rutile phase.

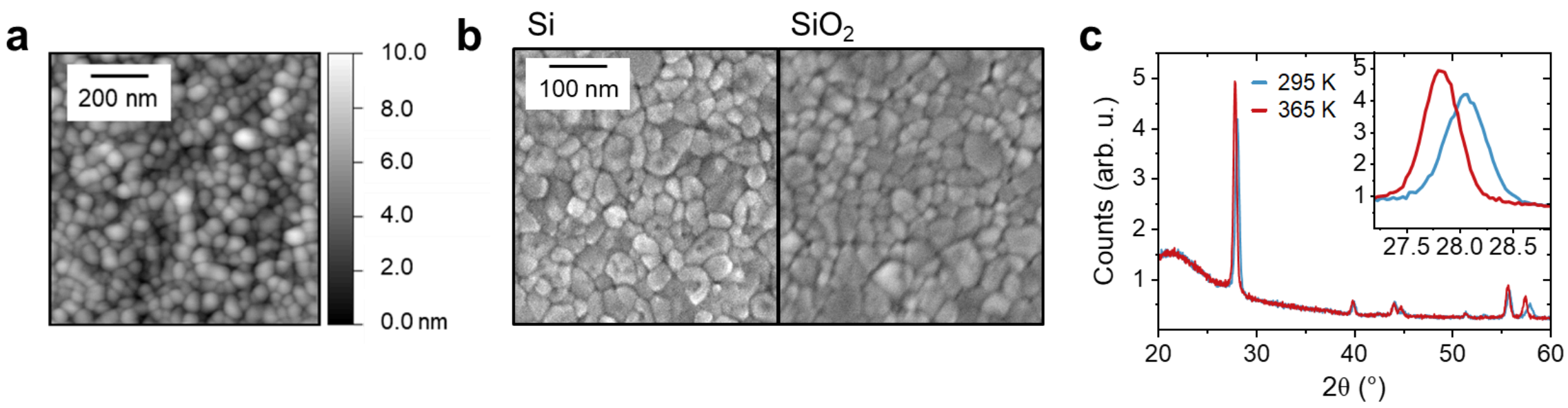

**Fig. S1. AFM and GI-XRD characterization of $VO_2$.** (a) AFM image of the 45 nm-thick $VO_2$ film on silica glass substrate. (b) SEM images of $VO_2$ films grown on silicon and silica substrates. (c) GI-XRD pattern below (295 K) and above (365 K) the transition temperature for the 45 nm-thick $VO_2$ film on silica glass substrate. The inset in (c) shows the (011) and (110) peaks of the insulating and metallic phase, respectively.

The temperature dependence of the optical properties was investigated by spectrophotometry. Fig. S2 shows the transmittance at a wavelength of 775 nm during heating and cooling cycles. This wavelength was selected because it provides a strong optical contrast between the insulating and metallic phases for both film thicknesses studied in this work.

The experimental data were fitted using a complementary error function[1] to extract key parameters of the phase transition, including the transition temperature $T_{\mathrm{IMT}}$ during both heating and cooling cycles. The fitting function is given by:

$$f_{\mathrm{erfc}}(T; T_{\mathrm{IMT}}, \sigma, A, y_0) = \frac{A}{2}\mathrm{erfc}\left(\frac{T - T_{\mathrm{IMT}}}{\sqrt{2}\sigma}\right) + y_0 \tag{S1}$$

The results of both the heating and cooling cycles for the 45 nm- and 180 nm-thick samples on silica are shown in Fig. S2 as dash-dotted lines. In the same figure, the transition temperature $T_{\mathrm{IMT}}$ and the broadening $\sigma$, which are the relevant parameters for the finite element simulations, are shown for both the branches. For the $VO_2$ film on silicon, the same transition temperature as that determined for the film on silica is assumed.

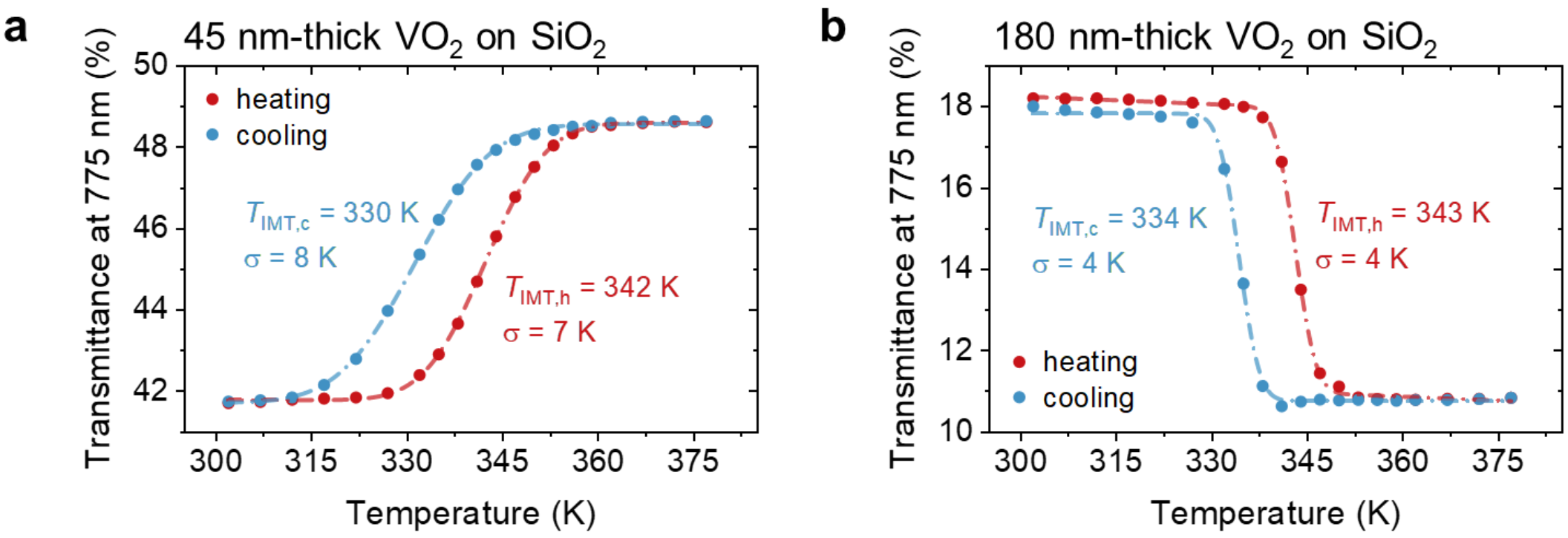

**Fig. S2. Transmittance hysteresis cycle for $VO_2$ on silica glass.** Temperature-dependent transmittance at 775 nm of $VO_2$ thin films with thickness (a) 45 nm and (b) 180 nm on silica substrate. The heating (red) and cooling (blue) experimental points are fitted with an erfc function (dash-dotted curves, Eq. (S1)) to extract the transition temperatures.

Finally, the experimental optical constants reported in Fig. 1 were used to calculate the reflectance spectra of all investigated samples through finite element simulations (see Methods section). The results, reported below, demonstrate that the reflectance strongly depends on both film thickness and substrate.

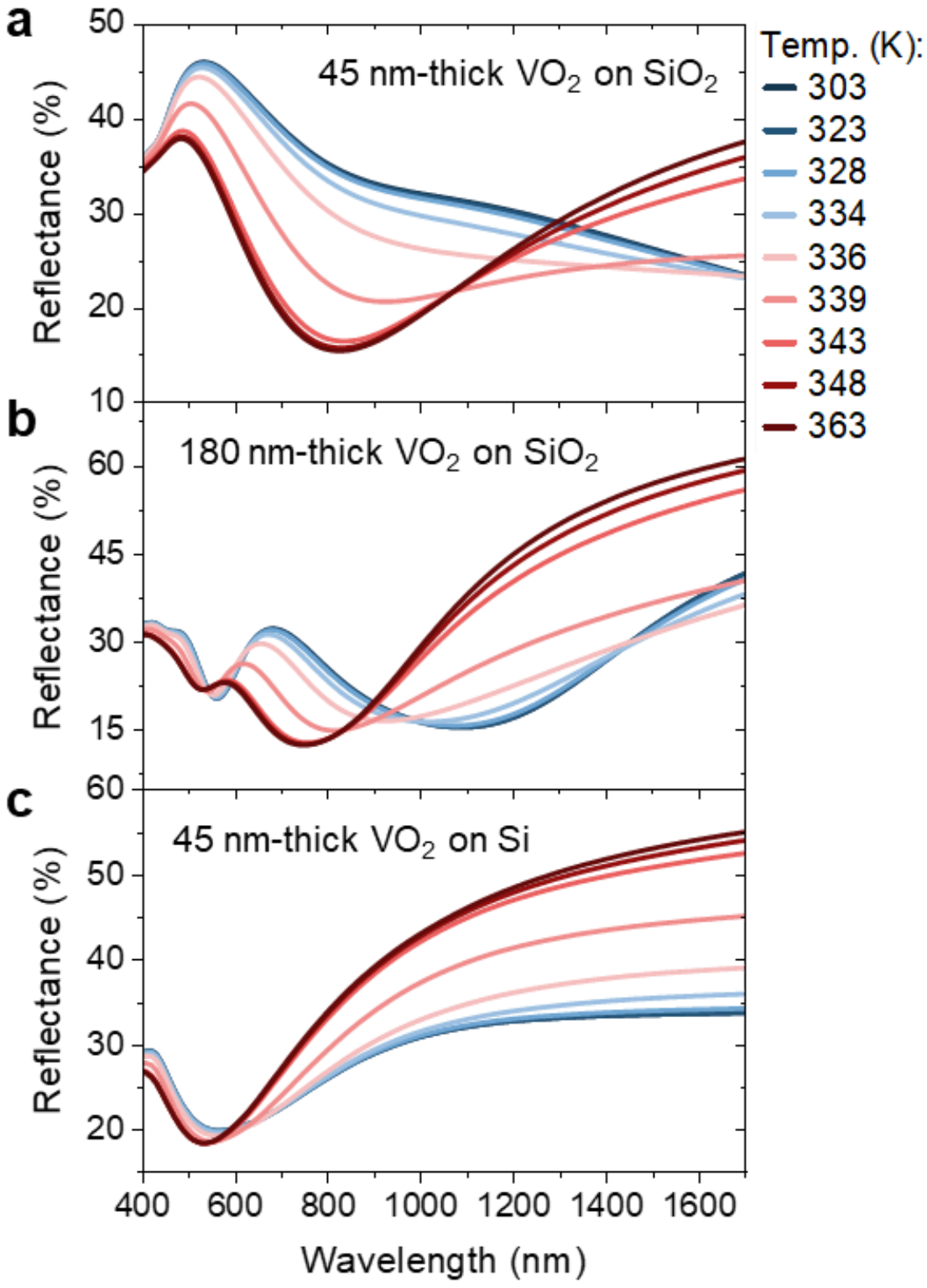

**Fig. S3. Simulated reflectance spectra at different temperatures for $VO_2$ thin films on silica and silicon substrate.** Temperature-dependent reflectance spectra of $VO_2$ thin films with thickness (a) 45 nm and (b) 180 nm on a silica substrate, and (c) with thickness 45 nm on silicon substrate.

## SUPPLEMENTARY SECTION 2:
## Spatiotemporal evolution of the reflectance variation

*Pump–probe images*: Representative stroboSCAT image time series for the three samples excited at 7.3 mJ/cm$^2$ (1/$e^2$ definition) are shown below. The radial dependence of the reflectance variation was extracted by performing an azimuthal average of each image and mirroring across the origin.

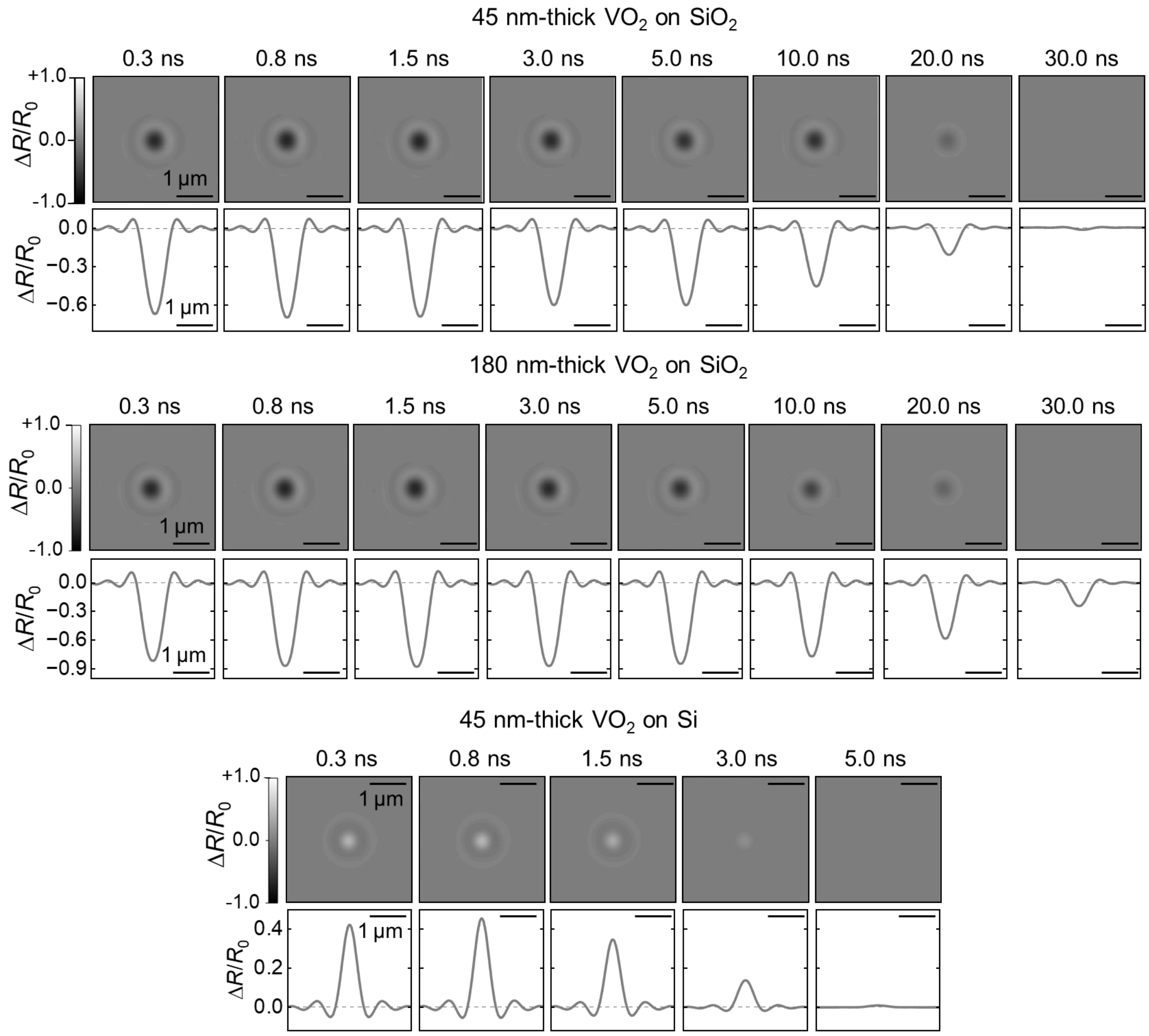

**Fig. S4. stroboSCAT time series and relative reflectance variation profiles for three $VO_2$ samples.** Time-resolved reflectance variation for 45 nm-thick sample on silica glass substrate, 180 nm-thick sample on silica glass, and 45 nm-thick sample on silicon. The representative data sets shown were photoexcited with 635 nm wavelength pump light pulses with a 1/$e^2$ fluence of 7.3 mJ/cm$^2$, and probed with 775 nm wavelength light pulses.

*Relaxation kinetics*: The relaxation kinetics of the photoexcited $VO_2$ thin films were investigated by tracking the decay of the peak relative reflectance change (normalized curves of the spatially integrated amplitude also overlap well with peak). Fig. S5 presents plots of relaxation curves that are supplementary to Fig. 3 of the main text, including data for the three samples for the range of excitation fluences studied. The rise of the signal (left-hand column) fits well to a Gaussian instrument response function (IRF) of 200 ps standard deviation. The normalized variations of the curves (right-hand column) highlight the differences in relaxation time. The thickness of the films and the substrate optical properties influence the saturation value and sign for the change in reflectance. The relaxation time scale is influenced by the film thickness, excitation fluence, and substrate thermal conductivity.

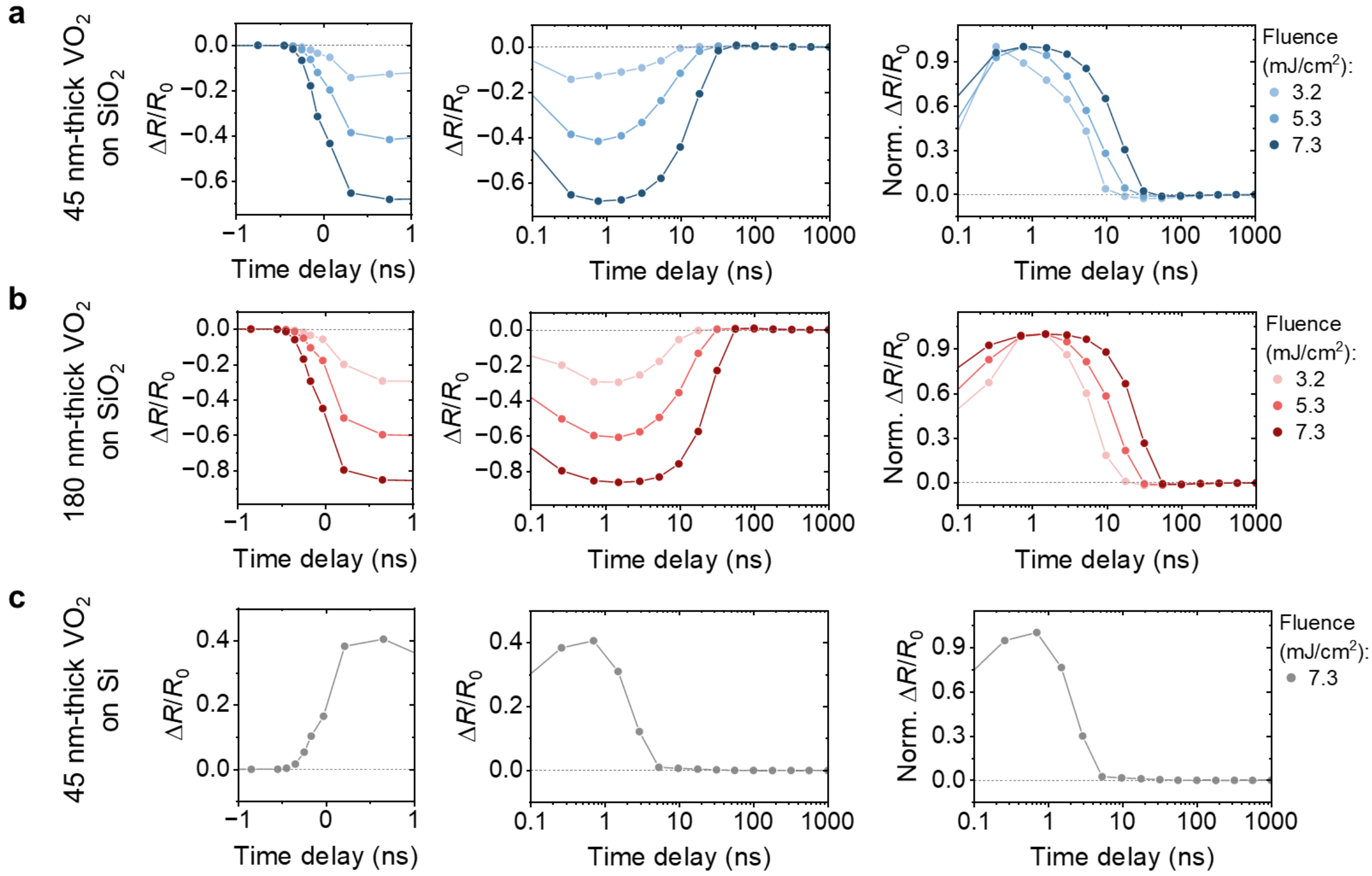

**Fig. S5. Relaxation dynamics of $VO_2$ on silica and silicon substrates.** Signal rise, peak reflectance variation time traces following photoexcitation, and traces normalized to their maximum absolute value amplitudes for (a) 45 nm- and (b) 180 nm-thick samples on silica glass, and (c) 45 nm-thick sample on silicon.

For each curve, the half-life values are reported in Table S1.

**Table S1. Half-life times of $VO_2$ on silica and silicon substrates.**

| Fluence ($mJ/cm^2$) | 45 nm on $SiO_2$ (ns) | 180 nm on $SiO_2$ (ns) | 45 nm on Si (ns) |
|---|---|---|---|
| 3.2 | 4 ± 1 | 5 ± 1 | – |
| 5.3 | 5 ± 1 | 11 ± 1 | – |
| 7.3 | 12 ± 2 | 22 ± 2 | 1.9 ± 0.5 |

## SUPPLEMENTARY SECTION 3:
## Optical scattering model and fitting procedure

A derivation of the optical scattering model implemented in this work is provided in the Methods Section. Fig. S6 presents a schematic of the multilayer geometry considered. Notably, the coverslip glass is omitted from the model owing to the use of index-matching oil. The air and oil layers are treated as semi-infinite media. The thicknesses ($L$) and optical constants ($n$, $k$) of the $VO_2$ thin films are experimentally determined by profilometry and ellipsometry (Fig. 1), respectively, and are reported in Fig. S6. Tabulated optical constants are used for the substrates, and the refractive index of the immersion oil is given by the manufacturer.

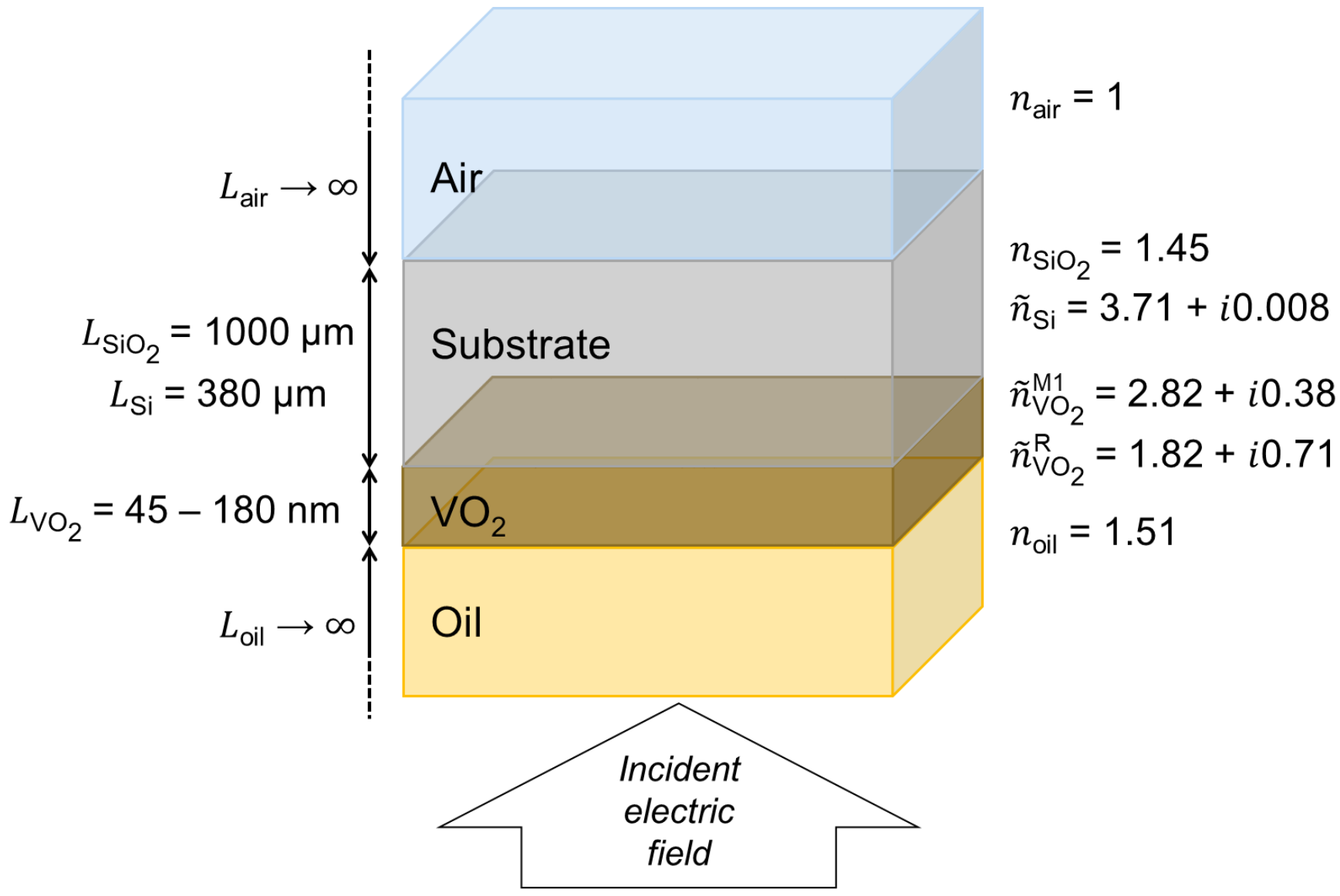

**Fig. S6. Scheme of the layers considered in the optical scattering model.** The four layers included in the derivation are shown together with their respective thicknesses and refractive indices at 775 nm.

The optical scattering model uses the spatial dependence of the complex refractive index in the $VO_2$ thin film. In the work of Ashoka *et al.*,[2] $\Delta\tilde{n}$ was modeled as a Gaussian radial profile and exponential attenuation in the $z$ direction, typical for small perturbations. In our analysis, the nonlinear optical properties of the phase transition in $VO_2$ leads to a spatial dependence of $\Delta n$ and $\Delta k$ that deviates significantly from a simple Gaussian form. Specifically, as shown in Fig. S7, if we assume a Gaussian radial distribution for the temperature within the films (Fig. S7a), which is uniform in the perpendicular direction, combined with the sigmoidal temperature dependence of $\Delta n$ and $\Delta k$ (Fig. S7b), yields radial profiles of the real and imaginary parts of the complex refractive index that exhibit a top-hat-like shape (Fig. S7c). The solid and dashed curves in Fig. S7b corresponds to fits obtained using a complementary error function to the ellipsometry data, analogous to the approach described earlier for the hysteresis cycle analysis (Eq. (S1)). The variations of the optical constants shown in Fig. S7c are parametrized using the expression in Eq. (1). It is important to note that the $z$-dependence of $\Delta\tilde{n}$ is neglected in our thin $VO_2$ films, since a few hundred picoseconds after excitation the temperature distribution evolution results in a nearly uniform spatial profile of $\Delta n$ and $\Delta k$.

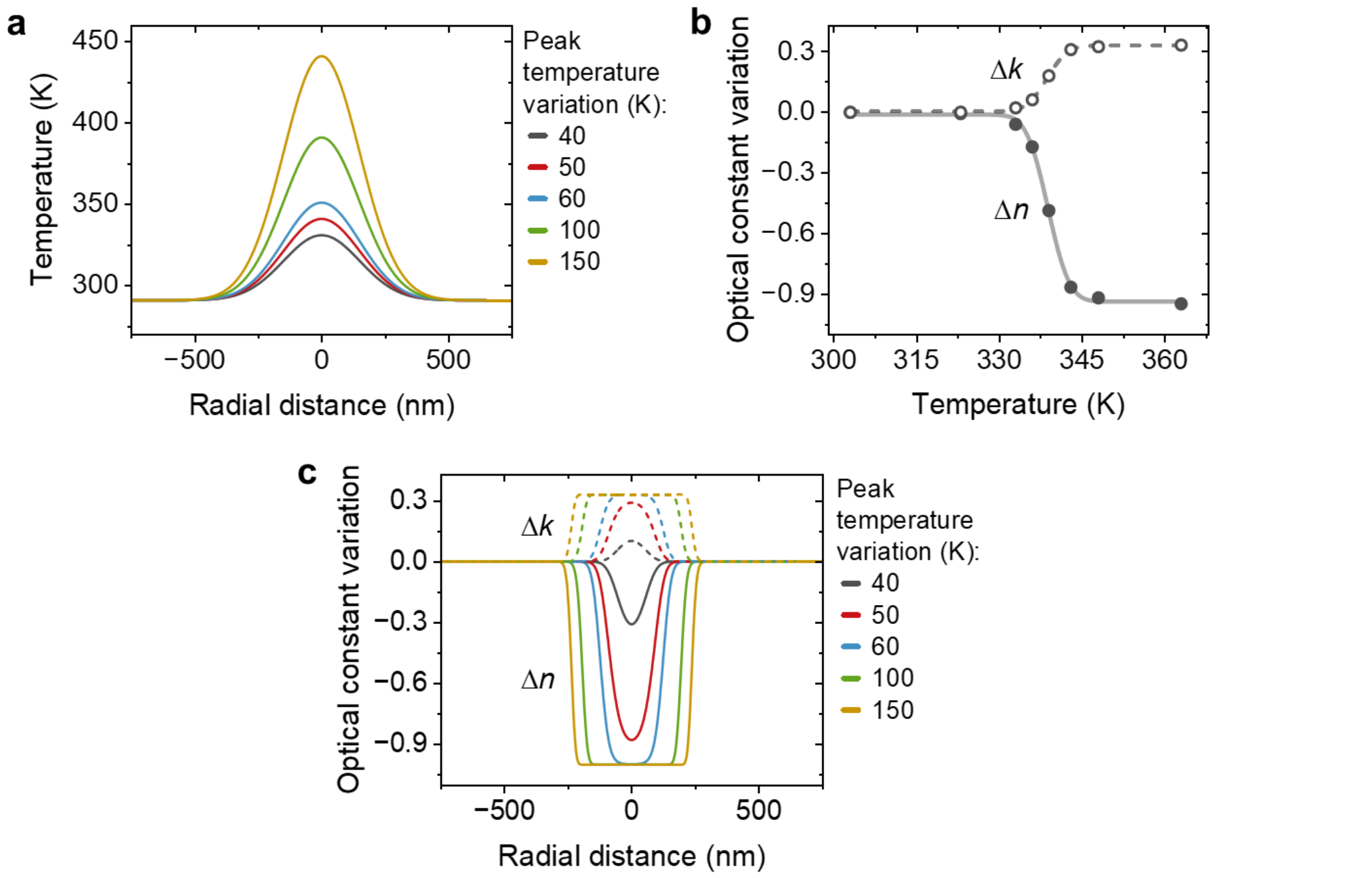

**Fig. S7. Representative calculation of the optical constant variation profile in a $VO_2$ thin film.** (a) Gaussian spatial profiles (350 nm fwhm) for representative peak temperatures inside the film. (b) Variation of the optical constants at 775 nm as a function of temperature. (c) Resulting spatial profile of the change in the optical constants at the same wavelength.

To verify that Eq. (1) is a good approximation to the expected refractive index profile throughout the entire evolution of broadened top-hat to Gaussian-like function, we directly fit profiles obtained from thermal transport simulations and ellipsometry data. Below we plot representative simulated temperature profiles for the 45 nm-thick $VO_2$ film on silica glass (from Fig. 4). To obtain the corresponding refractive index profile, we follow the procedure above, plugging the temperature profiles, $T(r)$, into the temperature-dependent refractive index curve of Eq. (S1) (Fig. 1c, S2), $\Delta n(T)$ and $\Delta k(T)$, to obtain the parametric curve $\Delta n(T(r))$ and $\Delta k(T(r))$. We fit the result to Eq. (1). Eq. (1) is an excellent fit to the expected refractive index profile for all times.

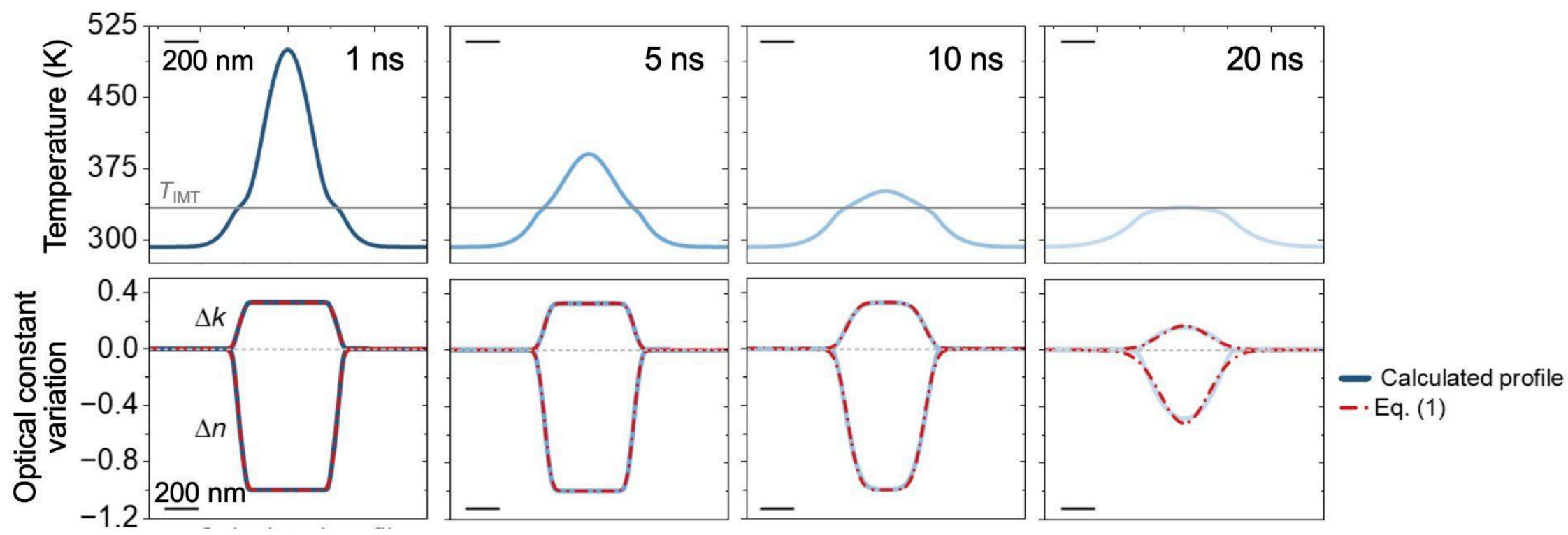

**Fig. S8. Fit of simulated refractive index change profile shapes.** Top row: Simulated photoinduced in-plane temperature profile of the 45 nm-thick $VO_2$ film on silica excited at 7.3 mJ/cm$^2$. The grey line indicates cooling transition temperature $T_{IMT,c}$. Bottom row: Refractive index change profile (blue solid lines) calculated parametrically by plugging temperature profile into Eq. (S1), assuming $\sigma$ = 8 K (Fig. S2). The profiles are fit to Eq. (1) (red dashed lines) with fixed $\Delta n$ = –1 and $\Delta k$ = –0.33, and only the width δ and the broadening ε floating. All scale bars are 200 nm.

Below we present representative fitting results of the optical model for three different underlying refractive index change profiles: Gaussian, top-hat, and Eq. (1) with variable broadening. At early times the Gaussian profile fit fails and the top-hat fits well, while in contrast at late times the Gaussian profile fits while the top-hat fails. Only Eq. (1) succeeds to fit the $\Delta R/R$ profile for all times by accounting for the smooth transition of the refractive index profile.

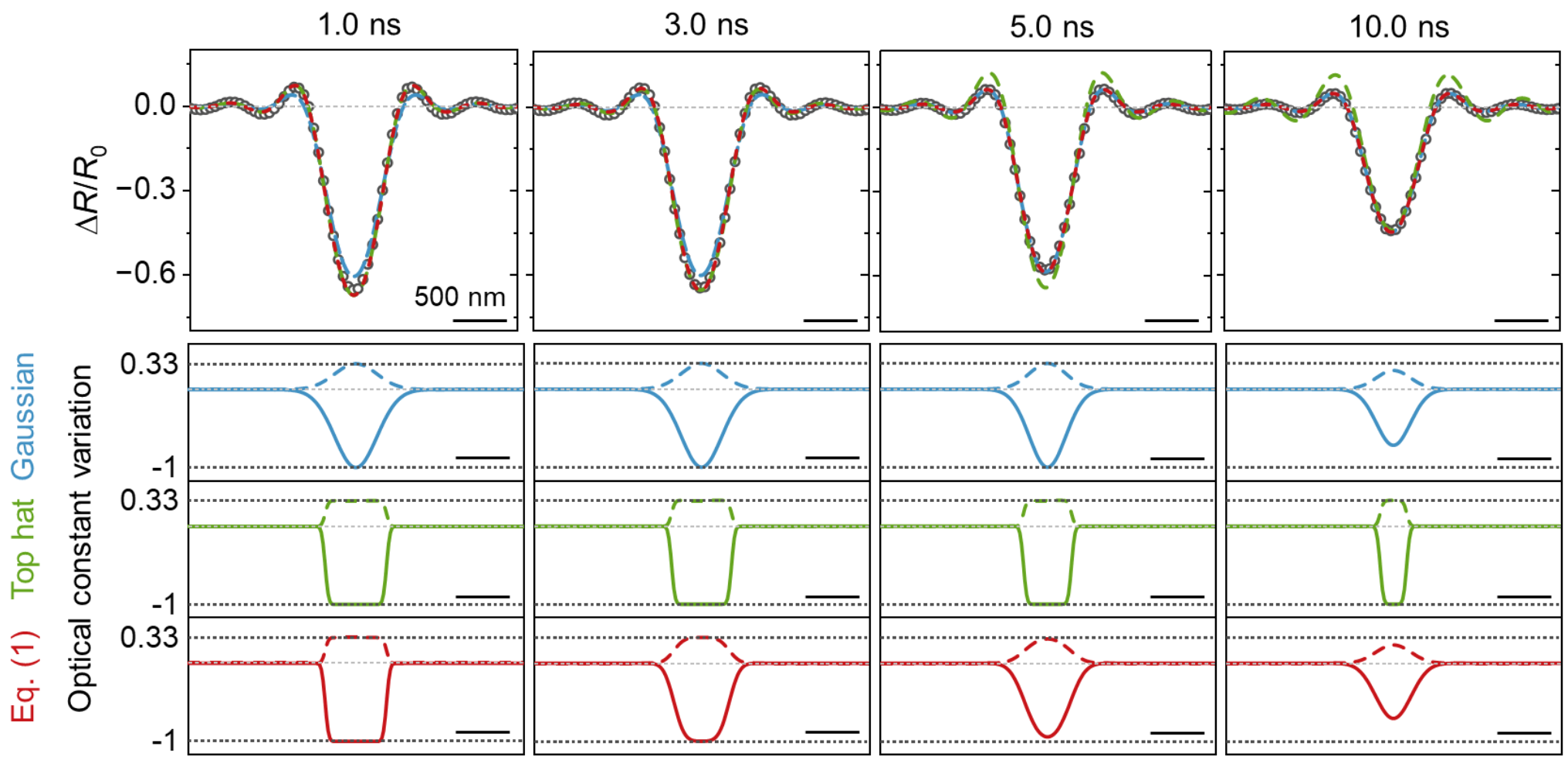

**Fig. S9. Comparison of fitting to optical scattering model for different refractive index change profile shapes.** Experimental data (points) for 45 nm-thick $VO_2$ film on silica glass excited with a fluence of 7.3 mJ/cm$^2$, and fitted curves (dashed lines) of the optical scattering model for the different refractive index profiles ($\Delta n$ as solid curve, $\Delta k$ as dashed curve) in the lower rows, including a Gaussian function (blue), a top-hat shape with fixed broadening (green), and Eq. (1) the broadened top-hat function with floating broadening term. All scale bars are 500 nm.

To experimentally verify the existence of the broadened top-hat profile for the change in reflectance, we performed steady-state thermoreflectance microscopy of the 45 nm-thick $VO_2$ sample on silica.[3] In the same home-built microscope as used for pump–probe measurements, we changed the 635 nm pump laser diode (≈350 nm fwhm) to continuous wave mode to induce a steady-state metallic domain. The steady-state reflectance image was collected using broadband lamp source spectrally filtered to a range of 700−750 nm, and the camera exposure time was 5 ms. As the cw pump power increases, the reflectance change relative to background, $\Delta R/R_0$, develops a clear top-hat-like shape, as expected from Fig. S7. This shape is more readily apparent in the steady-state version of the experiment because the laser diode is able to attain higher cw power than in pulsed mode, resulting in a metallic domain size exceeding the diffraction limit.

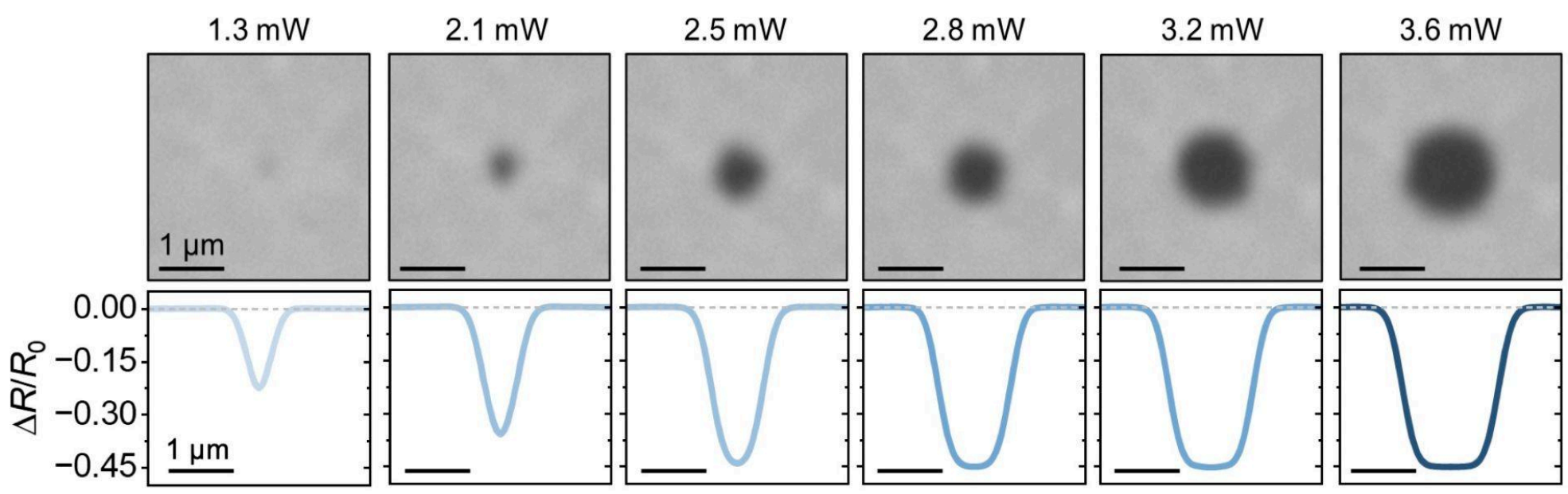

**Fig. S10. Steady-state thermoreflectance microscopy of $VO_2$.** Wide-field reflectance images as a function of power of a focused 635 nm wavelength cw laser beam (≈350 nm FWHM) for a 45 nm thick $VO_2$ thin film on silica. Top row: raw reflectance images. Bottom row: azimuthal average of data in top row, relative to background image, mirrored across the origin.

As presented in the Methods section, the adjustable parameters of the model (Eq. (1)) are: the amplitudes ($\Delta n$ and $\Delta k$), the width ($\delta$), and the broadening ($\varepsilon$) of the real and imaginary parts of the complex refractive index profile, and the defocusing term ($\Delta z$). The effect of each parameter on the relative change in reflectance at 775 nm and for a 45 nm-thick $VO_2$ film on silica is illustrated in Fig. S11. Notably, the variation in the optical constants primarily affects the peak amplitude in the interference pattern, whereas their widths and positions are governed by $\Delta z$, $\delta$, and $\varepsilon$.

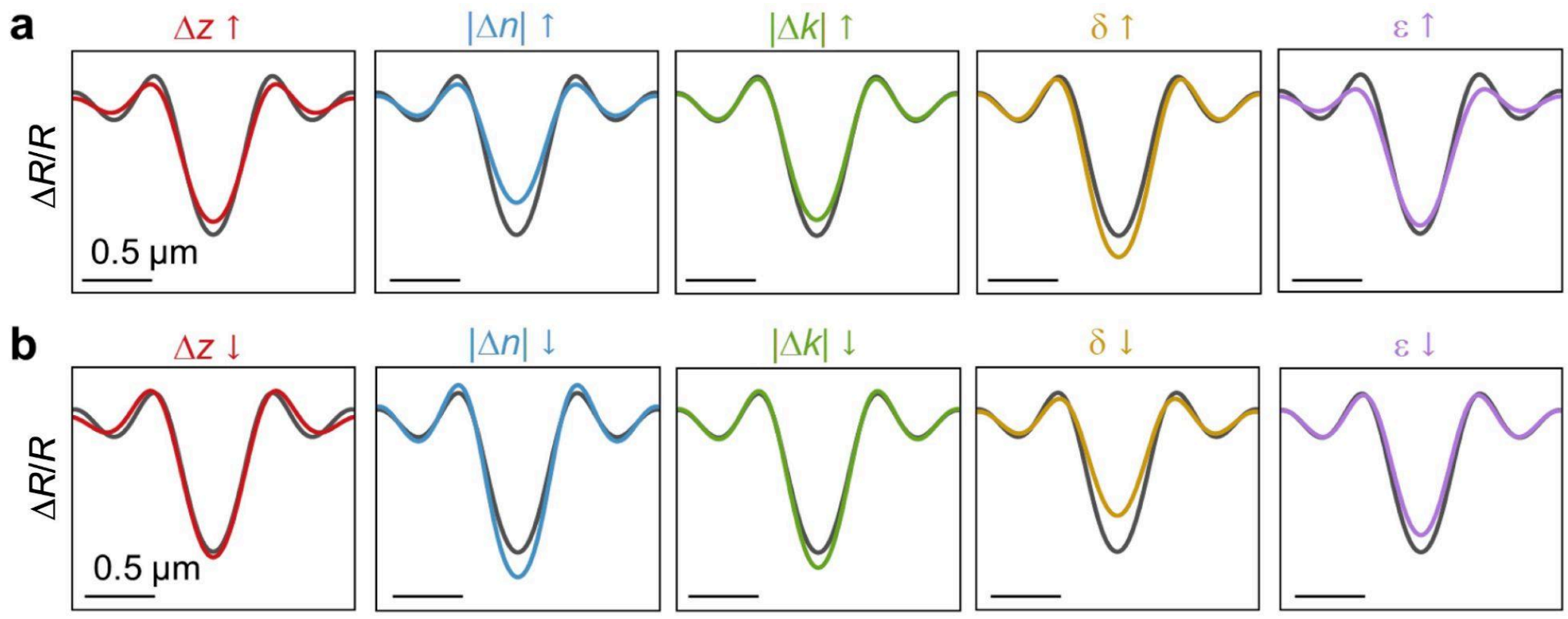

**Fig. S11. Effect of the different parameters of the optical scattering model.** Relative change in reflectance $\Delta R/R$ at 775 nm for 45-nm thick $VO_2$ film on silica substrate when $\Delta z$, $\Delta n$, $\Delta k$, $\delta$, and $\varepsilon$ are (a) increased or (b) reduced with respect to the black curves that correspond to the reflectance variation for $\Delta z = 0.0$ nm, $\Delta n = -1.0$, $\Delta k = -0.33$, $\delta = 130$ nm, and $\varepsilon = 50$ nm.

During the stroboSCAT measurements, the focal plane is kept fixed and the film is thin; thus, $\Delta z$ is not treated as a free parameter. Instead, its value—near $\Delta z \approx 0$ nm in our experiments—is determined through a global fit across all transient reflectance profiles. This fitting protocol yields values in the range of $|\Delta z| \leq 300$ nm, consistent with the uncertainty we estimate for focusing the microscope. Moreover, the real and the imaginary parts of the complex refractive index are kept constant and equal to the saturation values independently measured experimentally for our samples (Fig. 1d), i.e. $\Delta n = -1$ and $\Delta k = -0.33$. Thus, only the width $\delta$ and the broadening $\varepsilon$ are left free to vary. We note that allowing the optical constants to vary as well yields values in approximate agreement with the measured ones, but introduces unnecessary error into the fits.

The fitted stroboSCAT profiles and the corresponding profiles for $\Delta n$ and $\Delta k$ for some time delays are reported below for the three investigated samples. For every excitation fluence, there is a good agreement between the optical scattering model and the experimental data.

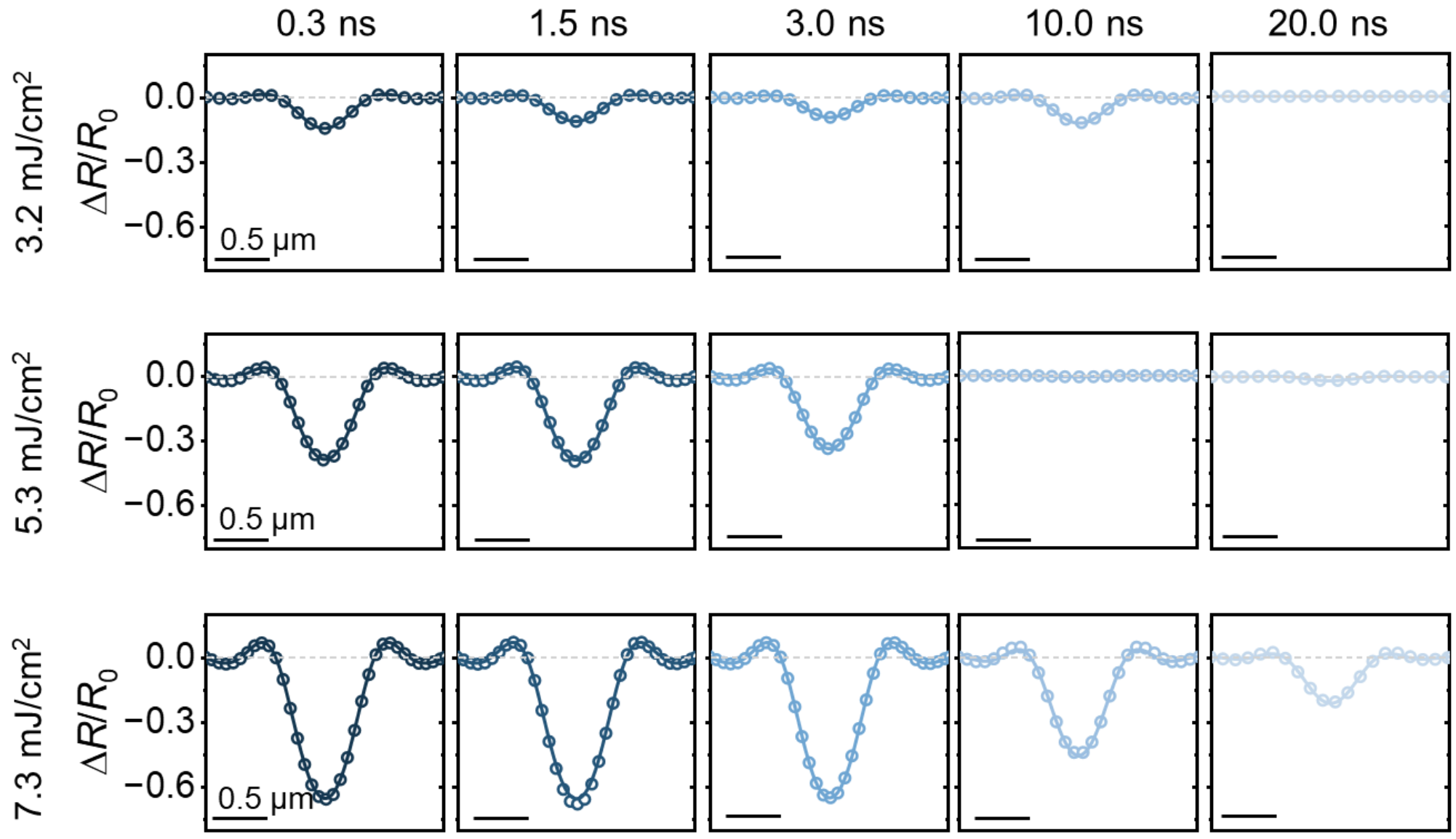

**Fig. S12. Fitting with optical scattering model for 45 nm $VO_2$ on silica glass.** Experimental data (points) and fitted curves (lines) for 45 nm-thick $VO_2$ film on silica glass for all the three investigated fluences at $\Delta z \approx 0$ nm.

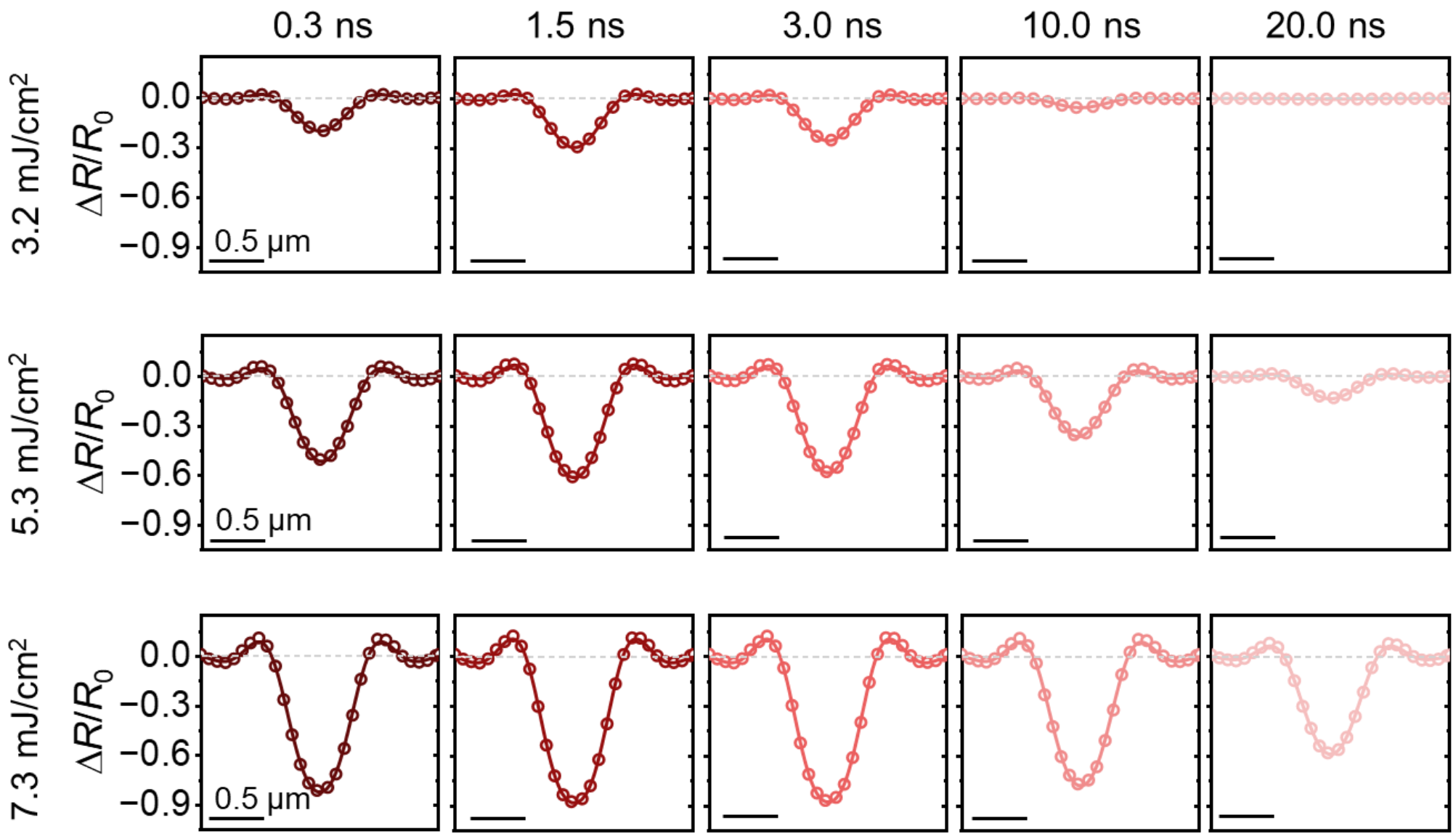

**Fig. S13. Fitting with optical scattering model for 180 nm $VO_2$ on silica glass.** Experimental data (points) and fitted curves (lines) for 180 nm-thick $VO_2$ film on silica glass for all the three investigated fluences at $\Delta z \approx 0$ nm.

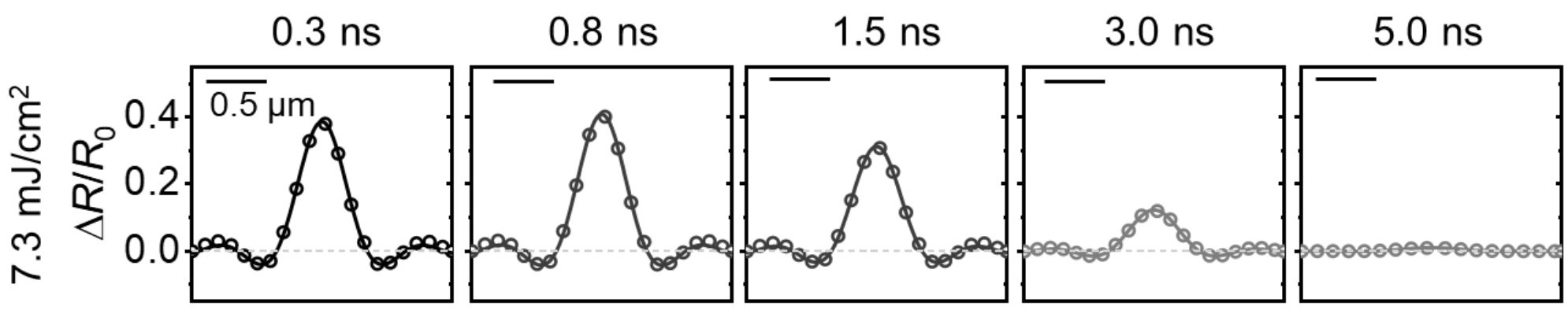

**Fig. S14. Fitting with optical scattering model for 45 nm $VO_2$ on silicon.** Experimental data (points) and fitted curves (lines) for 45 nm-thick $VO_2$ film on silicon at $\Delta z \approx 0$ nm.

The results of the fitting procedure are reported below for all the investigated samples.

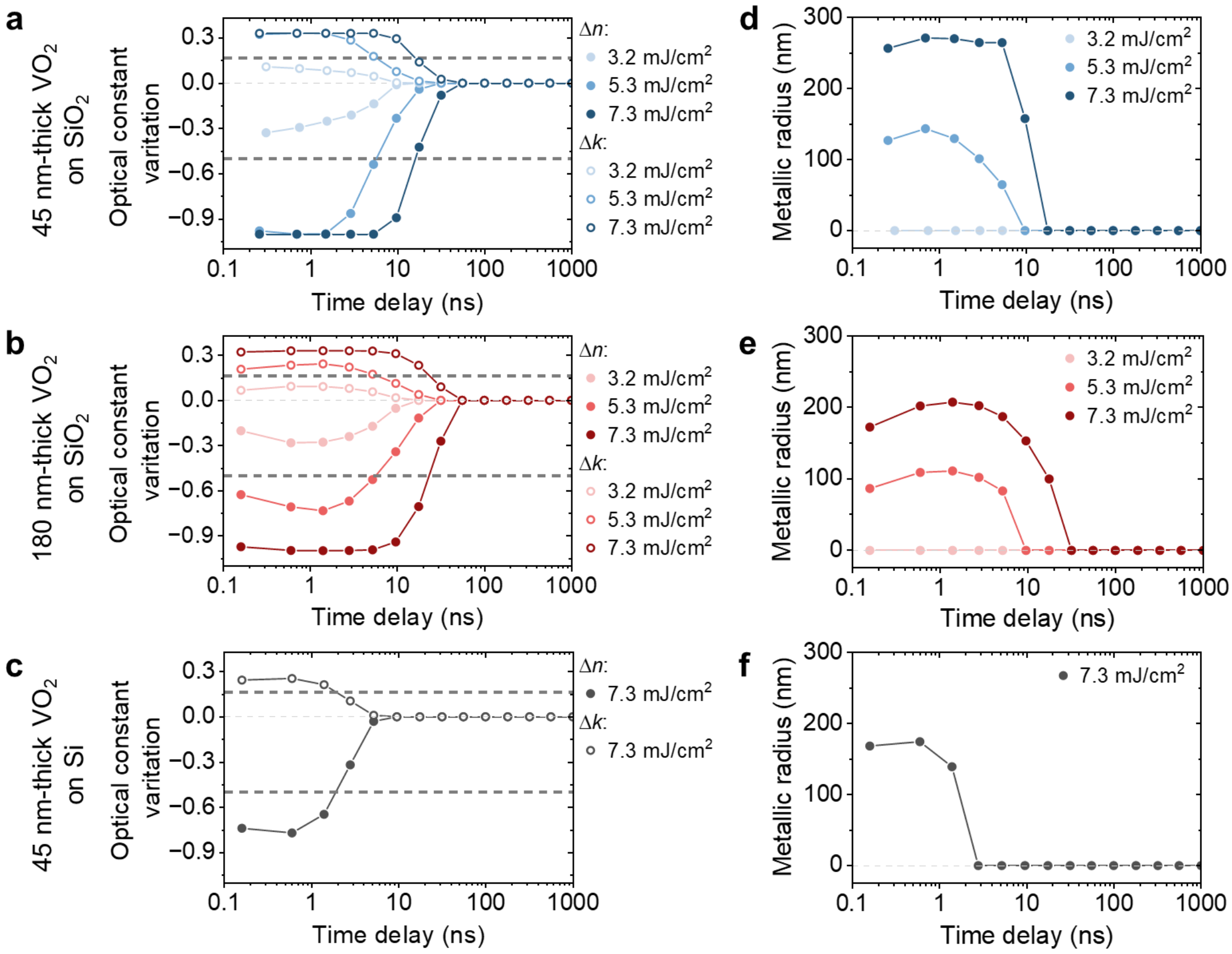

**Fig. S15. Results of the fitting with the optical scattering model for all samples.** (a-c) Optical constant variation at 775 nm at the center of the pump spot. The gray dashed lines indicate where the changes in the refractive index and extinction coefficient exceed half of their saturation values, corresponding to the onset of the insulator-to-metal transition. (e-f) Metallic domain radius evolution for the indicated investigated samples.

## SUPPLEMENTARY SECTION 4:
## Modulated thermoreflectance measurements

The thermal conductivity of a thin film can vary with thickness and fabrication method. Because thermal conductivity is key to our simulations and mechanistic interpretation, rather than using literature values we measured the thermal conductivities of our $VO_2$ thin films in both the insulating and metallic phases using modulated thermoreflectance microscopy. A detailed description of the experimental setup and analysis approach is presented in past works.[4–6] The $VO_2$ thin films were prepared on silicon substrates in the same way as the samples used for stroboSCAT measurements. We use a silicon substrate here to provide strong thermal contrast with the $VO_2$ layer. A 110 nm-thick Au transducer layer was evaporated on the $VO_2$ thin films to provide light-to-heat conversion and for thermoreflectance read-out. Measurements were performed on a heating stage with a scanning pump (532 nm, Cobolt MLD) and fixed probe (488 nm, Oxxius) configuration. The intensity of the pump beam was modulated by an acousto-optic modulator. The thermal conductivity of the transducer was found to be 80% the bulk Au value in a control measurement performed on a BK7 reference substrate using pump modulation frequencies of 100, 200, 400 and 800 kHz. For measurements of the samples with $VO_2$, 100 kHz was used as the modulation frequency. The beams were focused on the sample with an objective lens (N.A. = 0.5). The estimated diameter of both pump and probe were 2.5 µm. The maximum explored pump–probe distance was 15 µm, with a resolution of 1 µm and an acquisition time of 200 ms for each point. Two $VO_2$ thin films on silicon substrates were measured at a room temperature of 296 K and above the phase transition at 363 K. The experimental amplitude and phase profiles were fit according to numerical simulations based on Fourier's law to extract the thermal conductivity of the films.[7]

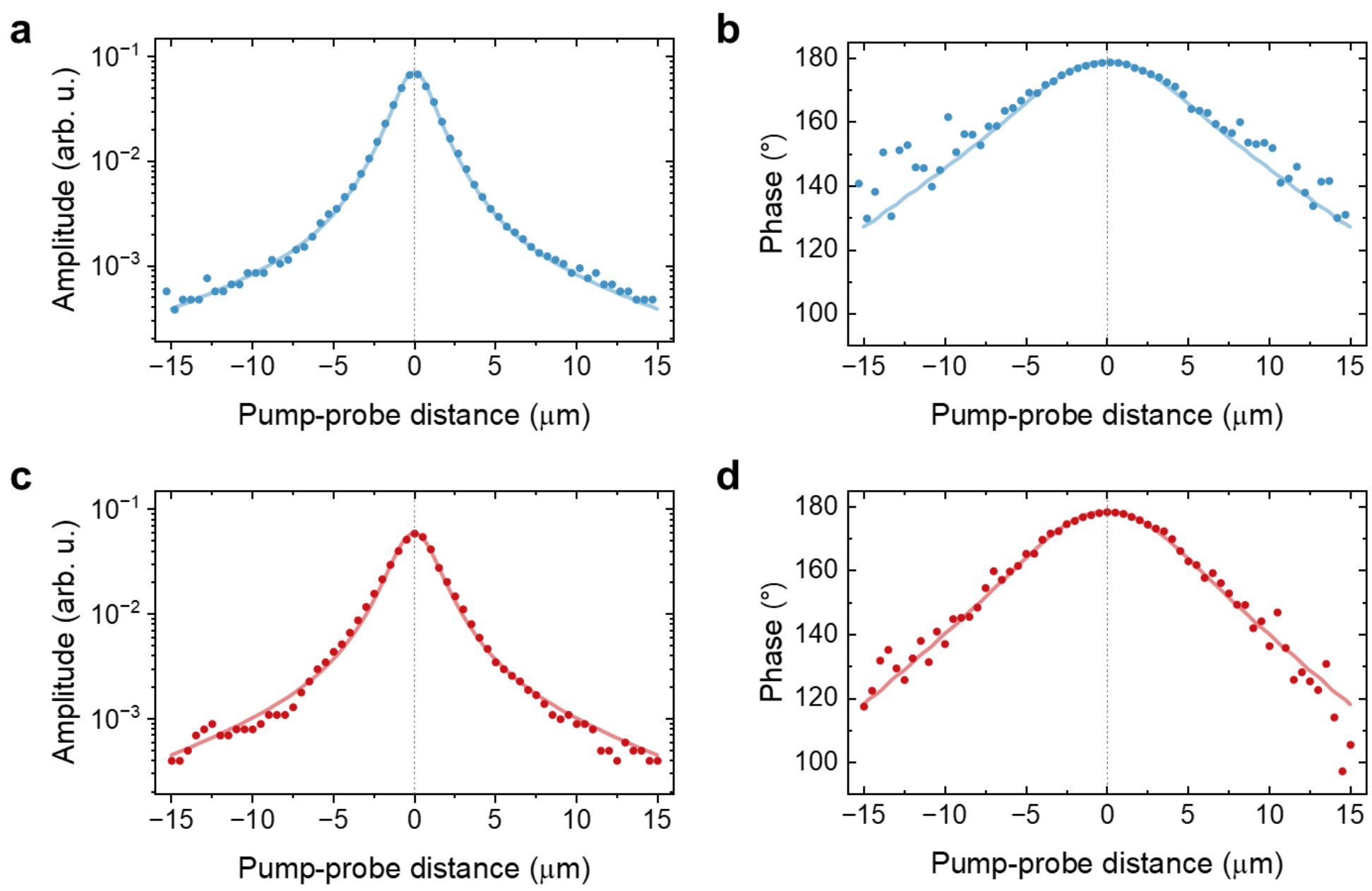

**Fig. S16. Amplitude and phase of the modulated thermoreflectance data for a 75 nm-thick $VO_2$ film.** (a,c) Amplitude and (b-d) phase for a 75 nm-thick $VO_2$ layer on silicon substrate and covered with 110 nm-thick Au transducer at (a,b) room temperature and (c,d) 90 ºC. The dots represent the experimental points, the solid lines correspond to the simulated values. The pump modulation frequency was 100 kHz.

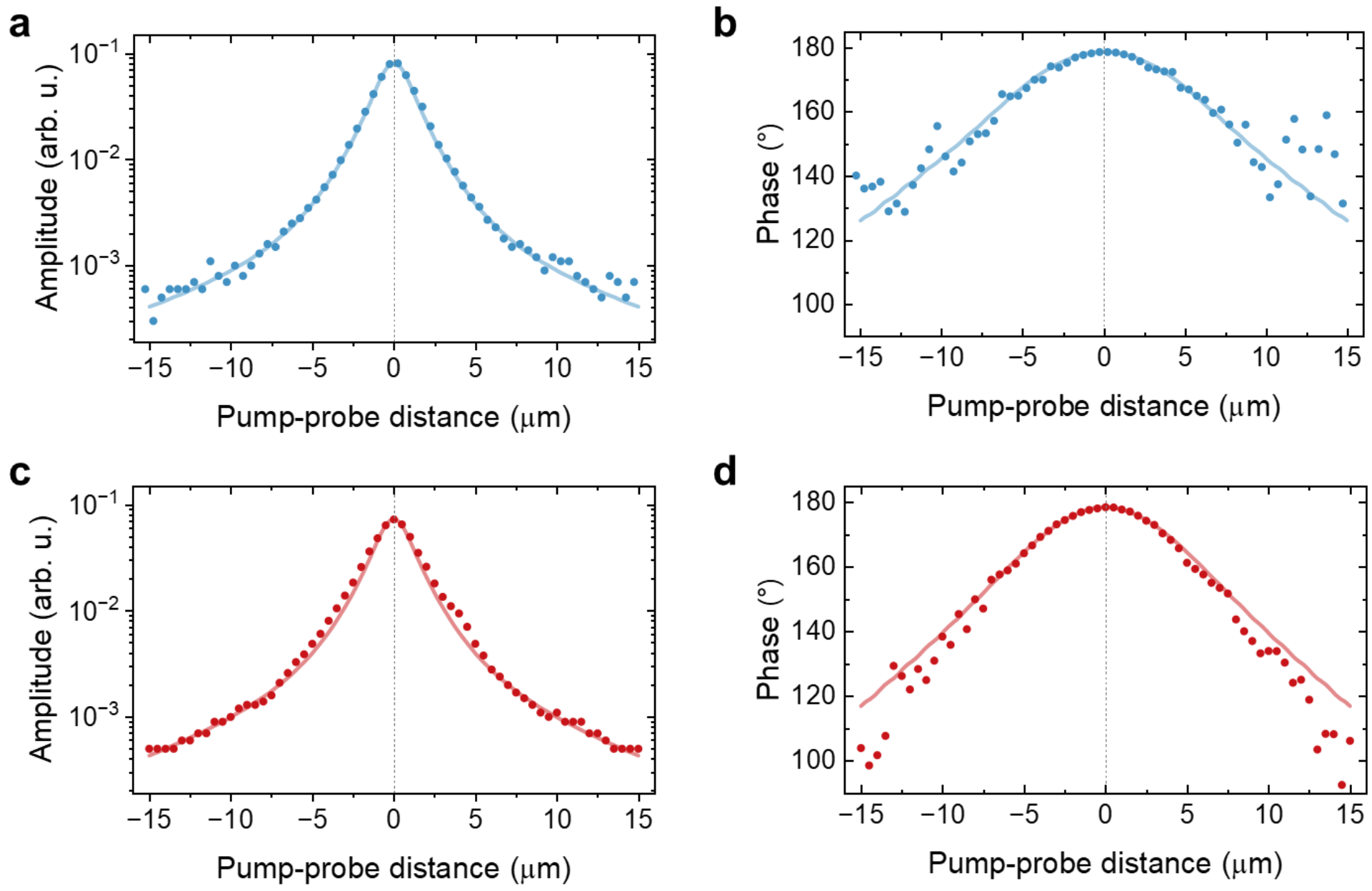

**Fig. S17. Amplitude and phase of the modulated thermoreflectance data for a 180 nm-thick $VO_2$ film.** (a,c) Amplitude and (b-d) phase for a 180 nm-thick $VO_2$ layer on silicon substrate and covered with 110 nm-thick Au transducer at (a,b) room temperature and (c,d) 90 ºC. The dots represent the experimental points, the solid lines correspond to the simulated values. The pump modulation frequency was 100 kHz.

From this analysis, we also extract the interfacial thermal resistance between $VO_2$ and the silicon substrate. An interfacial thermal resistance of the Au–$VO_2$ interface was not needed to obtain good fits. The interfacial resistance at the $VO_2$–Si interface was determined to be $1.0 \times 10^{-8}$ $m^2$ K $W^{-1}$. The experimentally determined values of thermal conductivity for both samples, in their insulating and metallic phases, are summarized in Table S2. These results are consistent with previous measurements of the thermal conductivity of polycrystalline $VO_2$,[8] with a small difference likely due to grain size. There, the observed increase in thermal conductivity going from the insulating to the metallic phase was attributed primarily to the introduction of the electronic contribution to heat conduction.

**Table S2. Experimental thermal conductivities of $VO_2$.** Thermal conductivities ($\kappa$) of the insulating monoclinic (M1) and metallic rutile (R) phases for $VO_2$ thin films on silicon substrate.

| Thickness | $\kappa_{M1}$ (W $m^{-1}$ $K^{-1}$) | $\kappa_R$ (W $m^{-1}$ $K^{-1}$) |
|---|---|---|
| 75 nm | 2.5 | 3.6 |
| 180 nm | 3.2 | 4.2 |

## SUPPLEMENTARY SECTION 5:
## Finite element simulations of the metallic radius evolution

*Temperature dependent thermal properties*: The light-to-heat conversion of the incident focused light pulse generates a transient temperature profile within the $VO_2$ thin film, which evolves according to Fourier's law of heat conduction. This process is simulated using the Heat Transfer in Solids module of COMSOL Multiphysics, accounting for both the sample geometry and substrate thermal effects. The three-dimensional problem can be reduced to a two-dimensional simulation by exploiting the radial symmetry of the studied system. The energy of a single Gaussian pump beam is converted into an incident power density by using the measured power, the duration (we assume 200 ps standard deviation for the simulation, approximately equal to the experimental IRF) and the spatial extent (350 nm full width at half maximum) of the pulse, the optical transmittance and reflectance of the samples, and specific heat capacity curve accounting for latent heat. Other experimental inputs that are considered are the heating and cooling transition temperatures obtained from transmittance hysteresis measurements (reported in Fig. S2), their broadening, as well as the thermal conductivities $\kappa$ of $VO_2$ in both phases as determined by modulated thermoreflectance (Table S2). For intermediate temperatures across the transition, the corresponding values of the thermal conductivity were interpolated from the hysteresis parameters, as illustrated in Fig. S18a. Modulated thermoreflectance technique also yields the interfacial resistance between $VO_2$ and the substrate, which influences heat dissipation (we assumed the same interfacial resistance at both $VO_2/SiO_2$ and $VO_2/Si$ interfaces). Additional parameters, including the density (we neglect the ≈1% change in density ($\rho$) between monoclinic and rutile phases of $VO_2$) and latent heat associated with the insulating-metal phase transition, were taken from literature.[9,10] In particular, the specific heat capacity ($c_p(T)$) exhibits a peak centered at the transition temperature, with an integrated area corresponding to the latent heat, $L$ = 51 J/g (Fig. S18b).[9,11,12] This causes a drastic decrease in the thermal diffusivity ($D(T) = \kappa(T)/\rho c_p(T)$) in proximity of the transition temperature, as shown in Fig. S18c, and this affects the expansion of the phase boundary. Finally, a uniform profile in depth is assumed for simplicity. This latter approximation is based on the ≈160 nm attenuation length of the pump light being larger than or comparable to the film thicknesses studied, partial reflection from the film–substrate interface that enhances absorption on a return pass, and finite-element simulations showing that any residual out-of-plane temperature gradient has largely equilibrated by the earliest experimentally accessible delay and moreover does not significantly impact the simulated domain size dynamics.

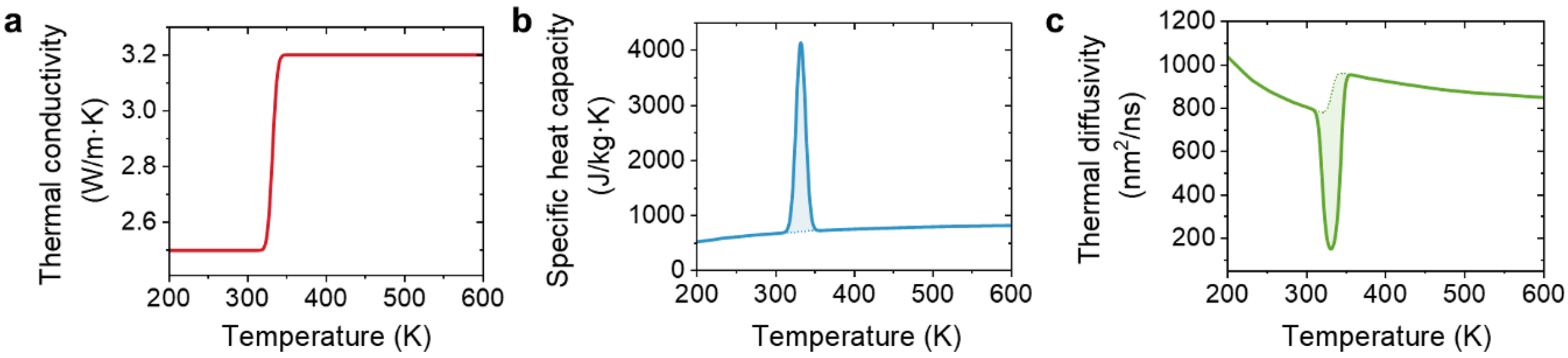

**Fig. S18. Input parameters for finite element simulations.** Temperature-dependent (a) thermal conductivity, (b) specific heat capacity, and (c) thermal diffusivity used in the finite element simulations for the 45 nm-thick sample. The transition temperature is $T_{IMT,c}$ = 330 K, while the shaded areas represent the effect of the latent heat on both the specific heat capacity and the thermal diffusivity.

*Simulated temperature profiles and metallic domain motio*n: Fig. S19a,b, S20a,b, S21a show the simulated radial temperature maps at selected time delays for the three samples. Time-zero is defined as the peak of the Gaussian pulse, reflecting the experiments. In the presence of the pump, the heating transition temperature is used, while for the subsequent thermal recovery the cooling transition temperature is employed.

From these temperature maps, the metallic radius can be extracted. The latter is evaluated as the radial distance at which the local temperature equals the experimental cooling transition temperature. To account for the finite width of the hysteresis branches in the polycrystalline film, $\sigma$, a range of transition temperatures is considered when evaluating its evolution and is visualized in the form of an uncertainty on the simulated metallic radius. Moreover, since the position of the transition isotherm varies slightly with depth in the film, we calculated the mean radial distance by averaging over the film thickness, yielding the effective metallic radius for a probe that integrates over the film thickness. The results of this procedure for all the investigated samples and fluences are shown in Fig. S19c,d, S20c,d, S21b.

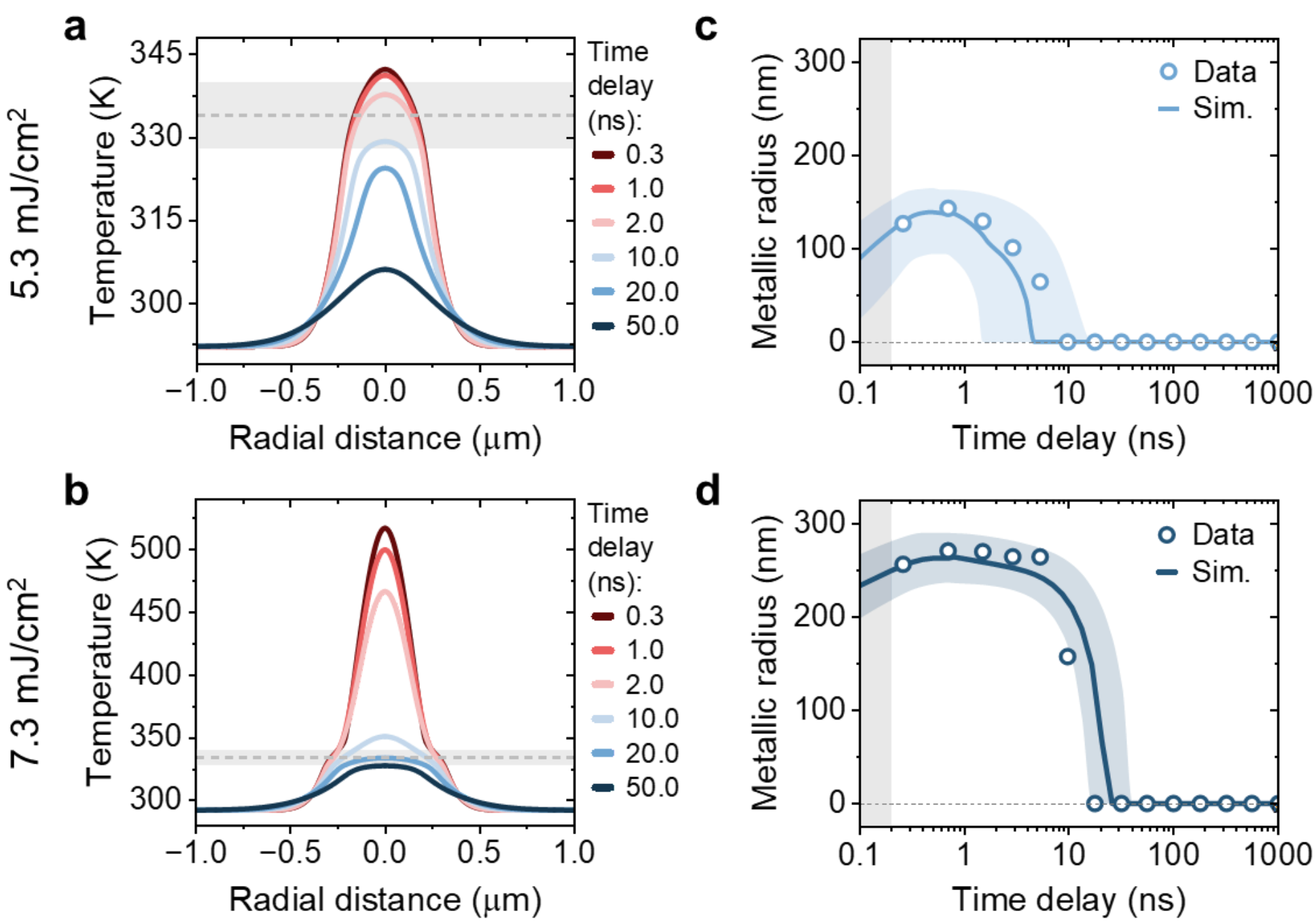

**Fig. S19. Evolution of the temperature profile and metallic radius for the 45 nm-thick $VO_2$ film on silica glass.** (a-b) Finite element simulations of the temperature profile evolution after excitation at (a) 5.3 mJ/cm$^2$ and (b) 7.3 mJ/cm$^2$. The dashed gray line represents the average transition temperature, while the shaded areas indicate the cooling transition temperature broadening, $T_{IMT,c} \pm \sigma$. (c-d) Evolution of the corresponding metallic domain radius excited at (c) 5.3 mJ/cm$^2$ and (d) 7.3 mJ/cm$^2$. The shaded areas correspond to the simulated metallic radius when considering transition broadening; for the 45 nm-thick film, $\sigma = 8$ K (Fig. S2).

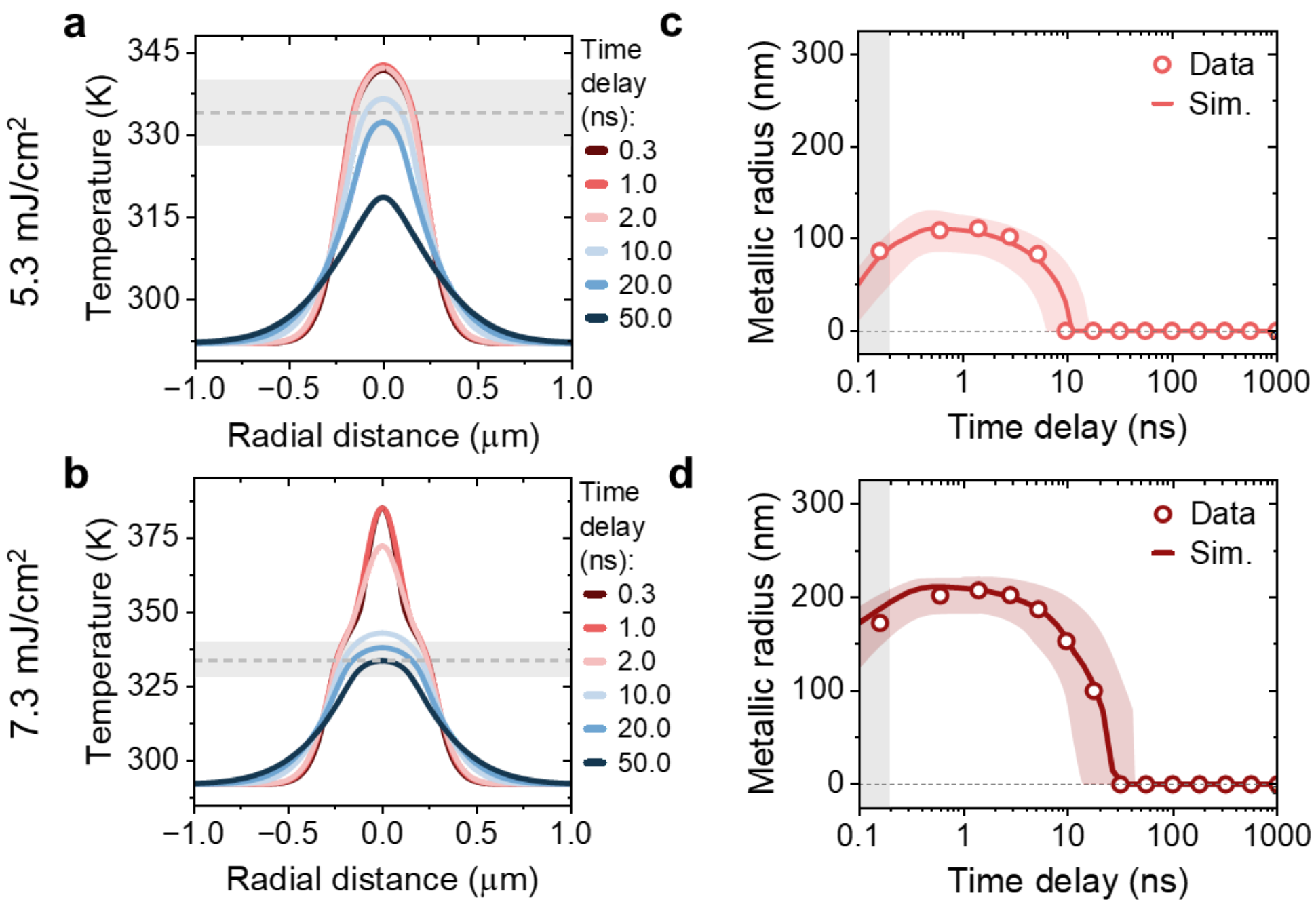

**Fig. S20. Evolution of the temperature profile and metallic radius for the 180 nm-thick $VO_2$ film on silica glass.** (a-b) Finite element simulations of the temperature profile evolution after excitation at (a) 5.3 mJ/cm$^2$ and (b) 7.3 mJ/cm$^2$. The dashed gray line represents the average transition temperature, while the shaded areas indicate the cooling transition temperature broadening, $T_{IMT,c} \pm \sigma$. (c-d) Evolution of the corresponding metallic domain radius excited at (c) 5.3 mJ/cm$^2$ and (d) 7.3 mJ/cm$^2$. The shaded areas correspond to the simulated metallic radius when considering transition broadening; for the 180 nm-thick film, $\sigma = 4$ K (Fig. S2).

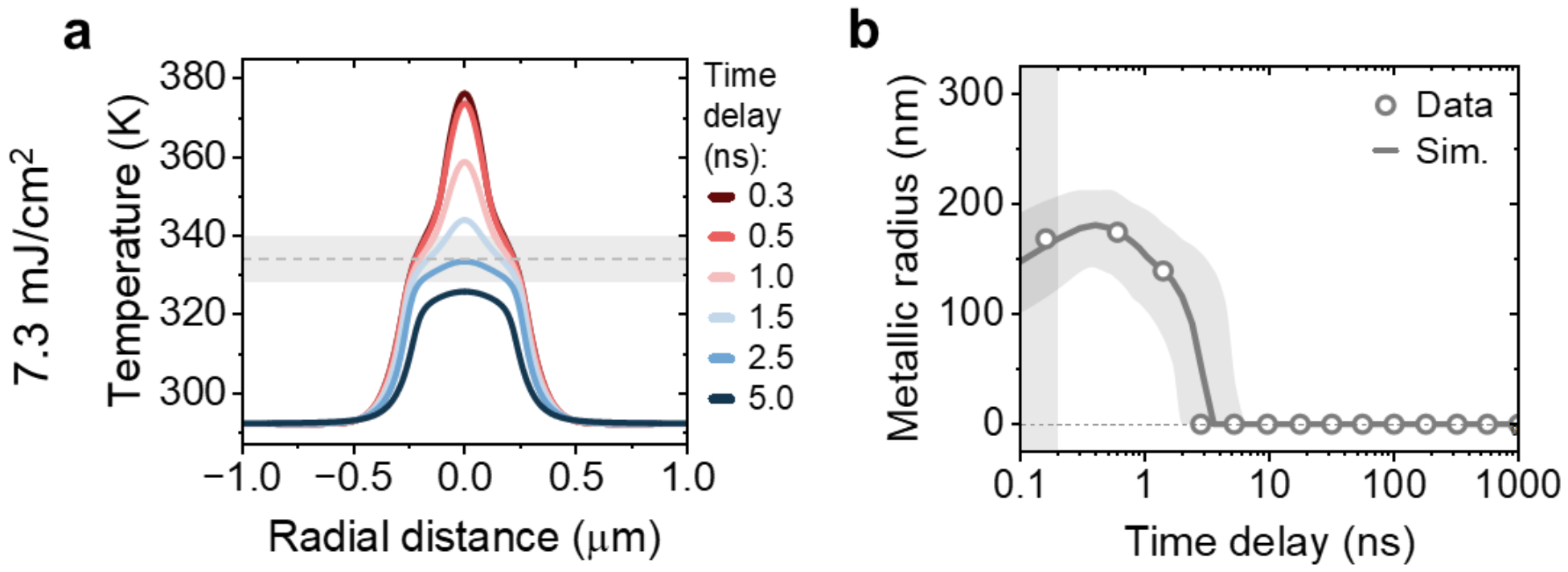

**Fig. S21. Evolution of the temperature profile and metallic radius for the 45 nm-thick $VO_2$ film on silicon.** (a) Finite element simulations of the temperature profile evolution after excitation at 7.3 mJ/cm$^2$. The dashed gray line represents the average transition temperature, while the shaded areas indicate the cooling transition temperature broadening, $T_{IMT,c} \pm \sigma$. (b) Evolution of the corresponding metallic domain radius excited at 7.3 mJ/cm$^2$. The shaded areas correspond to the simulated metallic radius when considering transition broadening; for the 45 nm-thick film, $\sigma = 8$ K (Fig. S2).

*Initial temperature profile shape sensitivity analysis*: In our simulations, we have assumed that the effective heat pulse has a Gaussian functional form in space, reflecting the shape of the pump beam. We note that this model assumes that there is no transient, non-diffusive component to the transport that could change the shape of the heating profile, for example, due to fast electronic transport occurring before lattice thermalization or by non-diffusive heat transport independent of thermalization. This assumption is based on the strongly scattering nature expected for the polycrystalline thin films that should lead to diffusive transport on the spatiotemporal scales accessed by the experiment. Still, we evaluate the possible impact of such a scenario. Given the tens of nm grain size and the spatiotemporal resolution of the experiment, such a process would most likely have consequences on the shape of the effective initial temperature profile (as opposed to non-diffusive transport on a nanosecond time scale). A sensitivity analysis appears below. Specifically, we performed finite element simulations of the phase front propagation, driven by thermal transport subject to latent heat, for a series of selected initial temperature profile functional forms. The chosen initial temperature profiles include: a simple Gaussian shape (as if the temperature profile had the same shape as the laser pulse and only broadened due to diffusion), a profile with shoulders around the transition temperature due to latent heat when applying Gaussian heating profile (the same profile shape generated by finite element simulations for all other simulations in this work), a central Gaussian with $1/r^2$ tails (reminiscent of a Lévy flight mechanism), and an exponential profile (representing ballistic transport subject to scattering). We find that while such a process may affect the initial width and thereby the quantitative details of the phase front propagation, the essential metallic domain size evolution is not strongly sensitive to the initial temperature profile shape.

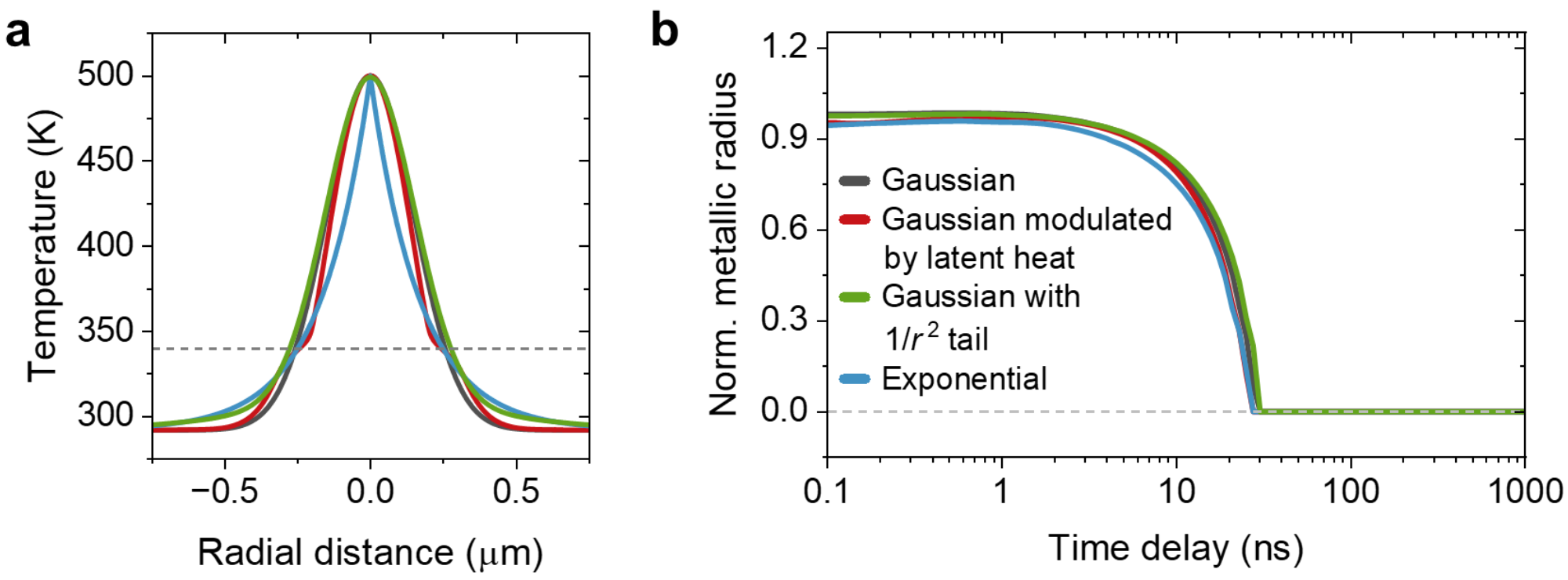

**Fig. S22. Sensitivity analysis of metallic domain size evolution to different initial temperature profile shapes.** (a) Representative initial temperature profiles used for finite element simulations of thermal transport. The chosen initial temperature profiles include a perfect Gaussian shape (black), a profile with shoulders around the transition temperature due to latent heat when applying Gaussian heating profile (red), a central Gaussian with $1/r^2$ tails (green), and an exponential profile (blue). The dashed gray line represents the transition temperature. (b) Finite element simulations of the evolution of the metallic domain radius, driven by thermal transport accounting for the latent heat of the transition, for the corresponding initial temperature profiles in panel (a).

## SUPPLEMENTARY SECTION 6:
## Simple model for phase boundary propagation faster than diffusion

Here we use a simple model to illustrate the point that transport of a spatial location corresponding to a fixed temperature along a thermal gradient is not limited to the thermal diffusivity of the overall profile. In this illustrative case, we neglect latent heat, instead only treating thermal diffusion with constant diffusivity independent of temperature. This model also neglects heat dissipation to the substrate.

In the following, all temperatures are relative to room temperature. Consider a radially symmetric two-dimensional Gaussian temperature gradient along the $r$ axis with initial peak temperature $\Delta T_{\mathrm{peak}}$ (relative to ambient temperature) and initial standard deviation $\sigma_0$ that evolves by diffusion with diffusivity $D$ as a function of time $t$:

$$\Delta T(r,t) = \Delta T_{\mathrm{peak}} 2\pi\sigma_0^2 \exp[-r^2/2(\sigma_0^2 + 2Dt)]/2\pi(\sigma_0^2 + 2Dt). \tag{S2}$$

Above we have maintained the factors of 2π that come from normalization factors to show where the expression comes from, but below we cancel them to give simplified forms. We are interested in finding the time dependence of the position $r = r_0(t)$ along the curve corresponding to a given temperature $\Delta T_0 = \Delta T(r_0)$ (relative to ambient temperature). Solving for $r_0^2$ gives

$$r_0^2(t) = 2\ln\{\Delta T_{\mathrm{peak}}\sigma_0^2/\Delta T_0(\sigma_0^2 + 2Dt)\}(\sigma_0^2 + 2Dt). \tag{S3}$$

We focus on $r_0^2(t)$ to connect to the mean-squared expansion that is the natural choice for diffusive transport. The metallic domain radius $r_0(t)$ is simply the square root of the above expression. This expression describes the time dependence of the location corresponding to the chosen temperature.

First, take the typical case where one tracks the amplitude corresponding to the $1\sigma$ width of the profile as it expands. The temperature at $1\sigma$, found by plugging in $r_0(t) = \sigma(t) = \mathrm{sqrt}[\sigma_0^2 + 2Dt]$, is $\Delta T_0(t) = \Delta T_{\mathrm{peak}} e^{-1/2}\sigma_0^2/(\sigma_0^2 + 2Dt)$. Plugging this in, we recover the well known expression for the mean-squared displacement of a Gaussian profile undergoing diffusion: $r_0^2(t) = \sigma_0^2 + 2Dt$.

Now, if instead one examines the expansion at $2\sigma$, one finds $r_0^2(t) = 4(\sigma_0^2 + 2Dt)$. That is, the position along the Gaussian temperature gradient corresponding to $2\sigma$ expands at an apparent rate that is four times faster than the standard deviation of the overall profile. This continues, giving a prefactor of $N^2$ when examining $N\sigma$. This is simply a mathematical consequence from looking at the expansion rate at a different position of the Gaussian curve; however, if that amplitude corresponds to the location of a physical process, then that enhanced expansion is physical.

In our case, $\Delta T_0$ would correspond to the temperature of the phase transition, which corresponds to the radius of the metallic domain at thermal equilibrium. This occurs at a fixed temperature independent of the evolving width of the profile: $\Delta T_0(t) = \Delta T_0$. In this case, the squared position is given by the full expression $r_0^2(t) = 2\ln\{\Delta T_{\mathrm{peak}}\sigma_0^2/\Delta T_0(\sigma_0^2 + 2Dt)\}(\sigma_0^2 + 2Dt)$. The time derivative of the position squared is:

$$\mathrm{d}/\mathrm{d}t(r_0^2(t)) = (\ln\{\Delta T_{\mathrm{peak}}\sigma_0^2/\Delta T_0(\sigma_0^2 + 2Dt)\} - 1)4D. \tag{S4}$$

Therefore, the instantaneous effective diffusivity of the phase boundary position is:

$$D_{\mathrm{eff}}(t) = [\mathrm{d}/\mathrm{d}t(r_0^2(t))]/2 = (\ln\{\Delta T_{\mathrm{peak}}\sigma_0^2/\Delta T_0(\sigma_0^2 + 2Dt)\} - 1)2D. \tag{S5}$$

Figure S23 illustrates the behavior predicted by this simple analytical model for a representative set of parameters.

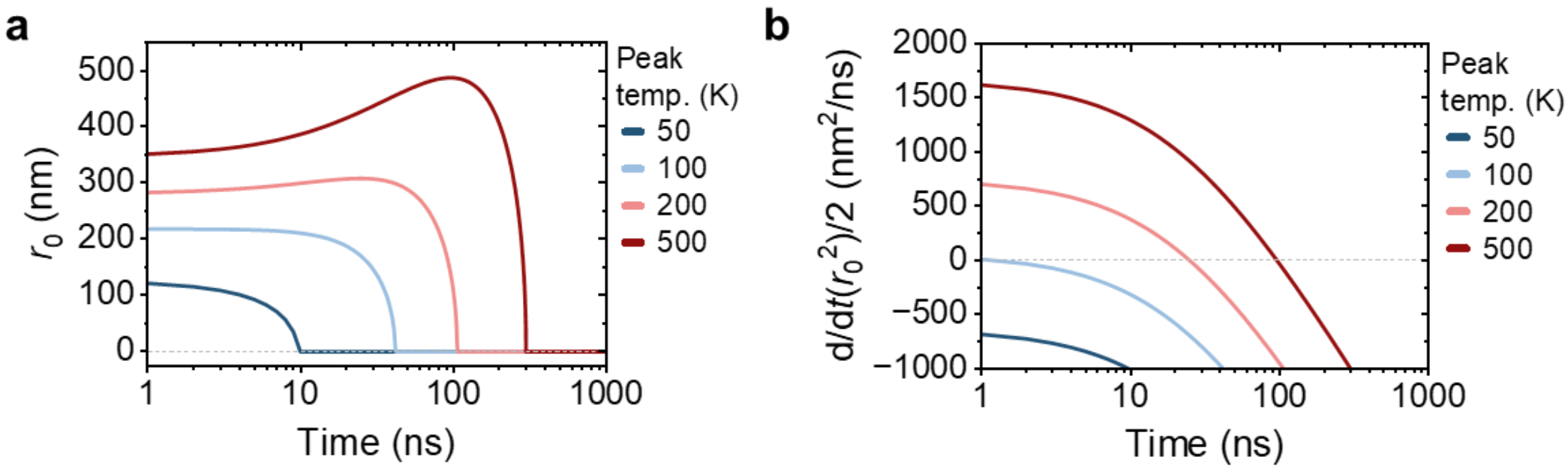

**Fig. S23. Simple diffusion model for the evolution of metallic domain radius and effective diffusivity.** Evolution of $r_0(t)$ (square root of Eq. S3) and $D_{\text{eff}}(t)$ (Eq. S5) for $\sigma_0 = 150$ nm, $D = 500$ nm$^2$/ns, $T_0 = 327$ K ($\Delta T_0 = 35$ K relative to ambient temperature of 292 K) for four selected $\Delta T_{\text{peak}}$ (relative to 292 K). Latent heat is not accounted for in this model.

Crucially, this simple model predicts that the apparent diffusivity of the phase boundary position is sensitive to the ratio $\Delta T_{\text{peak}}/\Delta T_0$ and gives both inward and outward propagation. The initial slope of this function provides intuition for the range of possible expansion rates. At early times, $t << \sigma_0^2/2D$, $\text{d/d}t(r_0^2(t)) \approx (\ln\{\Delta T_{\text{peak}}/\Delta T_0\} - 1)4D$. In particular, in the regime of initial outward propagation, the initial apparent diffusivity,

$$D_{0,\text{eff}}(t) = [\text{d/d}t(r_0^2(t))]/2 \approx (\ln\{\Delta T_{\text{peak}}/\Delta T_0\} - 1)2D \quad (\text{for } t << \sigma_0^2/2D), \tag{S6}$$

can exceed $D$ for $\Delta T_{\text{peak}} > e^{3/2}\Delta T_0$. According to the model, the initial transport increases logarithmically with the ratio $\Delta T_{\text{peak}}/\Delta T_0$; however, in reality the phase front will eventually fail to keep up with the isotherm.[13]

In the other limit, for small initial temperature rises in the range $\Delta T_0 < \Delta T_{\text{peak}} < e\Delta T_0$, the initial apparent diffusivity is negative, meaning the metallic domain would shrink from the beginning. As $\Delta T_{\text{peak}}$ approaches $\Delta T_0$ from above, the initial apparent diffusivity approaches a limit of $-2D$. Note that $\Delta T_{\text{peak}} < \Delta T_0$ is not defined as it corresponds to no phase change.

For the condition $\Delta T_{\text{peak}} = e\Delta T_0$, the initial diffusivity is zero. Including the impact of latent heat would further resist changes in the metallic domain size until it begins to collapse.

The apparent diffusivity is more generally a function of time, developing an increasingly negative slope in the $r_0(t)$ curve as time progresses up until the metallic domain disappears. Specifically, the metallic domain disappears ($r_0 = 0$) at a time $t_{\text{disappear}} = (\Delta T_{\text{peak}}/\Delta T_0 - 1)\sigma_0^2/2D$. The final apparent diffusivity attained just before the metallic domain fully shrinks is $D_{\text{eff}}(t_{\text{disappear}}) = -2D$, independent of the initial temperature rise.

This simplified model demonstrates the qualitative behavior observed in our simulations and experiments (compare Fig. S23 to Fig. 4 and 5), in particular the phenomenon of phase boundary transport that can be faster than thermal diffusion and its logarithmic dependence on initial temperature. It also provides insights about quantitative limits in the absence of latent heat. The true first-order transition material includes latent heat and nonlinear temperature dependence of thermal diffusivity that will modulate the above analysis, as will heat transfer to the substrate. We do not pursue an analytical model for those conditions, instead deferring to the finite element simulations (Fig. 5), which indeed show similar trends, though the latent heat attenuates the diffusivity.

## SUPPLEMENTARY SECTION 7:
## On instantaneous phase front diffusivity versus velocity

One of the important findings that emerges from our model for phase front propagation driven by thermal gradient evolution, is that instantaneous diffusivity is a more natural parameter than instantaneous velocity. The phase front velocity has been estimated under uniform excitation conditions, but only as a phenomenological parameter without connection to the underlying latent heat. Here, the mesoscale thermal dynamics model, validated by its agreement with the experimental data, indicates that diffusion is a more robust description of the transport in the present regime.

To compare the intrinsic instantaneous diffusivity and velocity, we take the metallic phase domain dynamics for a uniform out-of-plane profile and no substrate (Fig. 24a, reproduced from Fig. 5b) and calculate the two quantities. The instantaneous diffusivity was defined as $d/dt(r^2_{metal}(t))/2$, based on the mean-squared expansion description of diffusive transport. The instantaneous velocity was defined as $d/dt(r_{metal}(t))$. Fig. 24 shows the instantaneous diffusivity and velocity curves for the system with and without latent heat, for different initial temperatures.

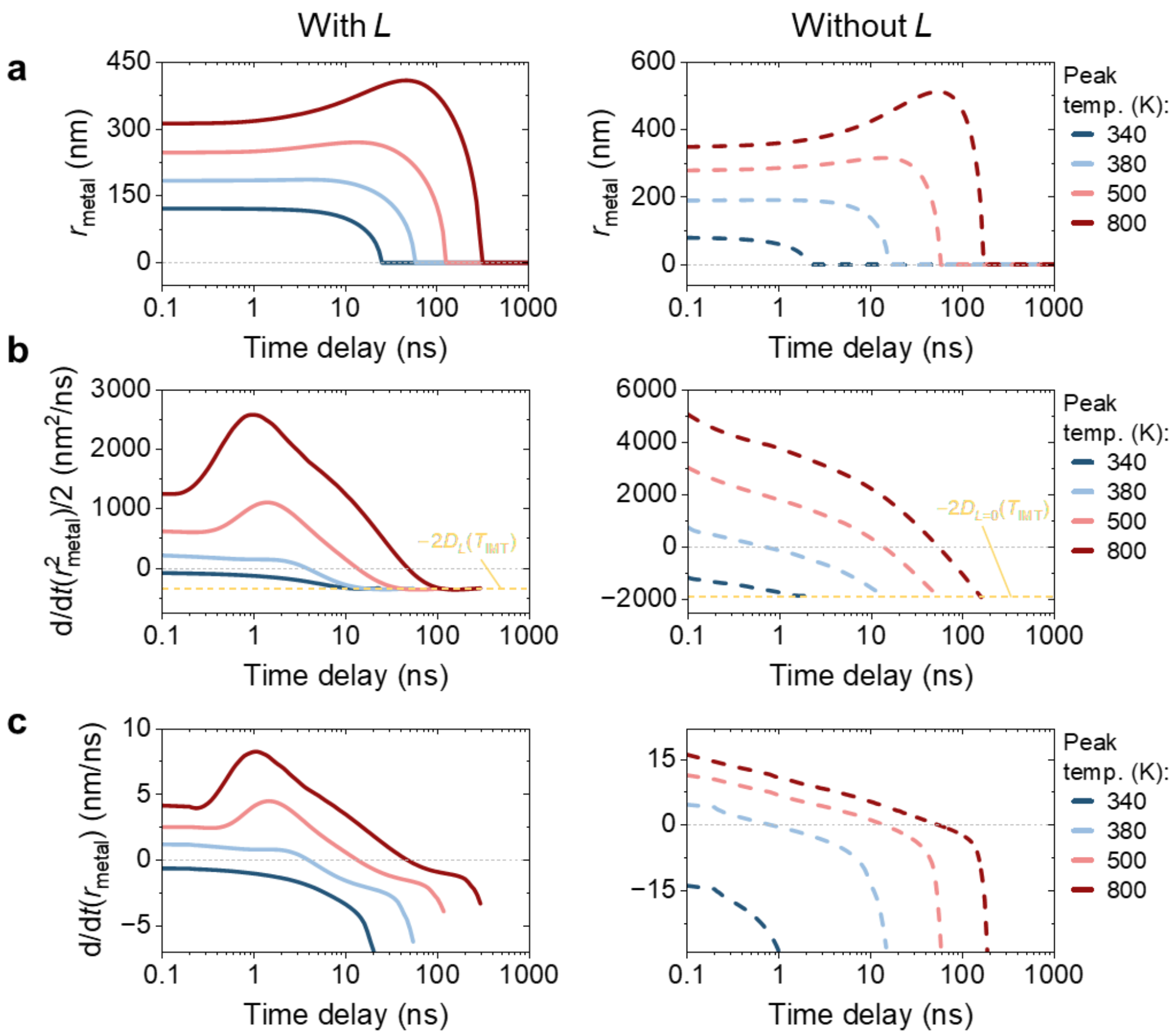

**Fig. S24. Comparison of instantaneous diffusivity and velocity.** (a) Simulated metallic radius normalized to the initial value for different peak temperatures, $r_{metal}(t)/r_{metal}(0)$, taking a uniform out-of-plane profile and no substrate (reproduced from Fig. 5). The solid and dashed curves show the evolution with and without the latent heat, $L$, respectively. These curves are used to obtain the instantaneous diffusivity curves, defined as $d/dt(r^2_{metal}(t))/2$, and instantaneous velocity curves, defined as $d/dt(r_{metal}(t))$. (b) Instantaneous diffusivity with and without latent heat. (c) Instantaneous velocity curves with and without latent heat. In all panels, the sign encodes the propagation direction, radially outward or inward.

We highlight the observation that the instantaneous diffusivity tends towards recurring values. In the case of finite latent heat, the initial diffusivity is temporarily constant, and the final inward propagation diffusivity stabilizes some time before disappearance to the same value, $-2D_L(T_{\mathrm{IMT}})$, independent of the initial temperature rise. Without latent heat, the diffusivity steadily decreases, but attains a well-defined final value just before disappearance, $-2D_{L=0}(T_{\mathrm{IMT}})$, again independent of the initial temperature rise. In contrast, the instantaneous velocity varies continuously with no well-defined behavior at any time throughout the phase front trajectory. Thus, we conclude that, under the condition of phase boundary propagation driven by dynamic thermal gradient evolution, diffusivity is a more reliable descriptor of underlying physical picture, consistent with a mesoscale thermal transport model based on Fourier's law.

Finally, we note that the simple analytical model of Supplementary Section 6 for the case of temperature-independent thermal diffusivity offers the same conclusion as the above simulations. As noted in the model, the final apparent diffusivity attained just before the metallic domain fully shrinks is $D_{\mathrm{eff}}(t_{\mathrm{disappear}}) = -2D$, independent of the initial temperature rise. However, if one examines the instantaneous velocity, $\mathrm{d}/\mathrm{d}t(r_0(t))$, the value varies continuously and approaches $-\infty$ as $t$ approaches $t_{\mathrm{disappear}}$. The true velocity likely has a physical upper limit, but this illustrates that the velocity is not a fundamental quantity of the transport in the thermally-driven regime.

## SUPPLEMENTARY SECTION 8:
## Experimental instantaneous diffusivity

One of the key analyses of this work is the extraction of the instantaneous diffusivity of the phase boundary propagation over the course of its evolution after the arrival of the heat pulse. This quantity is obtained by taking the time derivative of the square of metallic domain radius, $D_{eff}(t) = d/dt(r^2_{metal}(t))/2$, in analogy with the classic mean-squared expansion picture of diffusive transport, $r^2_{metal}(t) - r^2_{metal}(t_0) = 2D_{eff}t$.

In the main text we focus on the simulated $D_{eff}(t)$ for multiple reasons, the first being because we could readily access the intrinsic substrate-free dynamics, the second being that taking the derivative of experimental data in general amplifies noise and is unreliable. Keeping in mind that the experimental instantaneous diffusivity is quantitatively unreliable and subject to error, below we present the results of this analysis for all the investigated samples and pump fluences (45 nm $VO_2$ on $SiO_2$ at 5.3 and 7.3 mJ/cm$^2$ from Fig. 4 and S19, 180 nm $VO_2$ on $SiO_2$ at 5.3 and 7.3 mJ/cm$^2$ from Fig. 4 and S20, and 45 nm $VO_2$ on Si at 7.3 mJ/cm$^2$ from Fig. S21). The derivative was calculated as the average of the slopes between the point and its two closest neighbors, so the uneven time spacing of the data can introduce further systematic error.

We find that the experimental instantaneous diffusivities are generally consistent with the simulations. They exhibit similar behavior, including the crossover from positive values (outward propagation) at early times to negative values (inward propagation) at late times. Perhaps most importantly, the experimental data show the constant diffusivity for the final stage of the inward transport, as predicted by the simulations. This latter point is a signature of latent heat modulating the phase boundary motion. While we again caution against a quantitative interpretation of the experimental diffusivities, the magnitudes also are largely consistent with the simulations.

Taken together, this agreement between the experimental and simulated instantaneous phase boundary diffusion supports the conclusions drawn from the simulations in the main text.

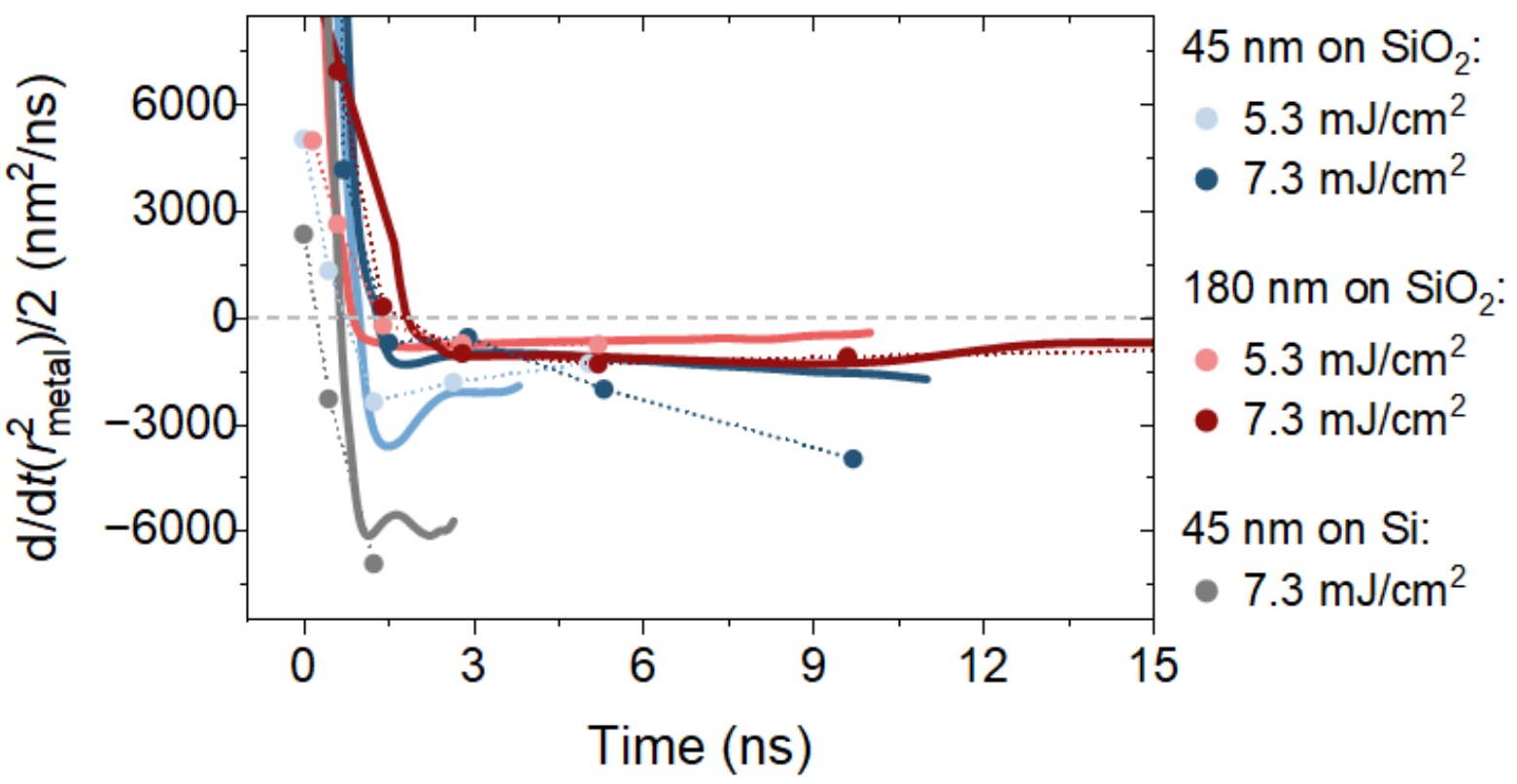

**Fig. S25. Comparison of experimental and simulated instantaneous diffusivity.** Instantaneous diffusivity versus time, $d/dt(r^2_{metal}(t))/2$, for experimental data (markers connected by dashed lines) and simulations (solid curves) for the indicated samples and pump fluence.

## SUPPLEMENTARY SECTION 9:
## Implications: Extended discussion

Our findings carry important implications for $VO_2$-based phase-change technologies. The ultrafast insulator–metal transition in $VO_2$ is attractive as a modulation mechanism for photonic applications; however, while the metallic phase can be induced on a sub-picosecond timescale, its relaxation is comparatively slow and not well controlled. It is thus important to identify the key control parameters for device speed and reproducibility. We showed that the recovery timescale, limited by thermal dissipation, can be tuned from tens of nanoseconds down to a few nanoseconds simply by controlling film thickness, substrate thermal conductivity, and excitation fluence. The intrinsic metal-to-insulator transition time may be even shorter, in which case the thermal relaxation time could be tuned over an even wider range. Thus, one could imagine achieving sub-nanosecond relaxation using the ambient temperature, engineering heat sinking, or material modification such as doping.

We anticipate that the strong refractive index change of the phase change combined with control over phase-front propagation will open opportunities for on-demand optical patterning and the design of photoinduced optical metamaterials. A major expected advantage of using $VO_2$ for such applications is that the nonlinear optical contrast of the metal–insulator transition presents sharp edges. Our results imply that the spatial extent of transient or steady-state metallic domains can be predicted by thermal transport modelling, and moreover controlled by film thickness, substrate thermal conductivity, and excitation fluence. We draw attention to the particular case where, under the right excitation conditions, the photoinduced metallic $VO_2$ domain remains spatially confined to its nucleation site until it collapses (Fig. 5, Supplemental Material Section 6). In this regime, the refractive index modification in turn remains spatially confined, ensuring that the written patterns would be well defined and reproducible, without being blurred by thermal or electronic diffusion. As a result, controlled nanostructuring of the refractive index could be achieved, enabling strategies for dynamic and reconfigurable photonic architectures.

## SUPPLEMENTARY REFERENCES